\documentclass[10pt,english,prd,superscriptaddress,nofootinbib,preprintnumbers,showpacs,floatfix]{revtex4-1}
\usepackage[utf8]{inputenc}
\usepackage{amsmath}
\usepackage{amsfonts}
\usepackage{amssymb}
\usepackage{graphicx}
\usepackage[usenames,dvipsnames]{xcolor}
\usepackage{hyperref}   
\hypersetup{colorlinks=true, 
citecolor=MidnightBlue, linkcolor=MidnightBlue, urlcolor=Cyan}
\usepackage{tikz}
\usepackage{verbatim} 
\definecolor{purple}{rgb}{1,0,1}
\definecolor{lime}{HTML}{A6CE39} 

\newcommand{\orcidicon}{%
	\begin{tikzpicture}
	\draw[lime, fill=lime] (0,0) 
		circle [radius=0.16] 
		node[white] {{\fontfamily{qag}\selectfont \tiny ID}};
	\draw[white, fill=white] (-0.0625,0.095) 
		circle [radius=0.007];
	\end{tikzpicture}	\hspace{-2mm}
}
\newcommand\orcidFrancisco{{\href{https://orcid.org/0000-0002-9388-8373}{\orcidicon}}}
\newcommand\orcidManuel{{\href{https://orcid.org/0000-0001-8586-0285}{\orcidicon}}}
\newcommand\orcidTarciso{{\href{https://orcid.org/0009-0007-0450-2672}{\orcidicon}}}
\begin{document}
\title{Black holes and regular black holes in coincident $f(\mathbb{Q},\mathbb{B}_Q)$ gravity coupled to nonlinear electrodynamics}


\author{Jos\'{e} Tarciso S. S. Junior\orcidTarciso\!\!}
	\email{tarcisojunior17@gmail.com}
\affiliation{Faculdade de F\'{\i}sica, Programa de P\'{o}s-Gradua\c{c}\~{a}o em 
F\'isica, Universidade Federal do 
 Par\'{a},  66075-110, Bel\'{e}m, Par\'{a}, Brazil}

	\author{Francisco S. N. Lobo\orcidFrancisco\!\!} 
	\email{fslobo@ciencias.ulisboa.pt}
\affiliation{Instituto de Astrof\'{i}sica e Ci\^{e}ncias do Espa\c{c}o, Faculdade de Ci\^{e}ncias da Universidade de Lisboa, Edifício C8, Campo Grande, P-1749-016 Lisbon, Portugal}
\affiliation{Departamento de F\'{i}sica, Faculdade de Ci\^{e}ncias da Universidade de Lisboa, Edif\'{i}cio C8, Campo Grande, P-1749-016 Lisbon, Portugal}

	\author{Manuel E. Rodrigues\orcidManuel\!\!}
	\email{esialg@gmail.com}
	\affiliation{Faculdade de F\'{\i}sica, Programa de P\'{o}s-Gradua\c{c}\~{a}o em 
F\'isica, Universidade Federal do 
 Par\'{a},  66075-110, Bel\'{e}m, Par\'{a}, Brazil}
\affiliation{Faculdade de Ci\^{e}ncias Exatas e Tecnologia, 
Universidade Federal do Par\'{a}\\
Campus Universit\'{a}rio de Abaetetuba, 68440-000, Abaetetuba, Par\'{a}, 
Brazil}

\date{\LaTeX-ed \today}
\begin{abstract}

In this work, we consider an extension of the symmetric teleparallel equivalent of General Relativity (STEGR), namely, $f(\mathbb{Q})$ gravity, by including a boundary term $\mathbb{B}_Q$, where  $\mathbb{Q}$ is the non-metricity scalar. 
More specifically, we explore static and spherically symmetric black hole and regular black hole solutions in $f(\mathbb{Q},\mathbb{B}_Q)$ gravity coupled to nonlinear electrodynamics (NLED).
In particular, to obtain black hole solutions, and in order to ensure that our solutions preserve Lorentz symmetry, we assume the following relation $f_Q = -f_B$, where $f_{Q}=\partial f/\partial\mathbb{Q}$ and $f_{B}= \partial f/\partial\mathbb{B}_Q$. We develop three models of black holes, and as the starting point for each case we consider the non-metricity scalar or the boundary term in such a way to obtain the metric functions $A(r)$. Additionally, we are able to express matter through analytical solutions for specific NLED Lagrangians ${\cal L}_{\rm NLED}(F)$. Furthermore, we also obtain generalized solutions of the Bardeen and Culetu types of regular black holes, by imposing specific metric functions.

\end{abstract}
\maketitle
\def\HMS{{\scriptscriptstyle{\rm HMS}}}
\bigskip
\hrule
\tableofcontents
\bigskip
\hrule


\section{Introduction}\label{sec1}

The discovery of the recent accelerated expansion of the Universe \cite{SupernovaSearchTeam:1998fmf,SupernovaCosmologyProject:1998vns} has spurred much interest in modified theories of gravities \cite{Capozziello:2011et,Clifton:2011jh,CANTATA:2021ktz}, as an alternative to dark energy \cite{Copeland:2006wr} and a possible cause of this late-time cosmic speed-up. In fact, a plethora of modified gravity theories have been explored in the literature that involve generalizations of the Hilbert-Einstein action, such as the $f(R)$ gravity \cite{Sotiriou:2008rp} and extensions \cite{Harko:2018ayt,Harko:2011kv,Bertolami:2007gv,Harko:2010mv,Harko:2011nh}, and the inclusion of second order curvature invariants such as $R_{\mu \nu }R^{\mu \nu }$, $R_{\alpha \beta \mu \nu }R^{\alpha \beta \mu \nu }$,  $C_{\alpha \beta \mu \nu }C^{{\alpha \beta \mu \nu}}$, $\varepsilon ^{\alpha\beta \mu \nu }R_{\alpha \beta \gamma \delta }R_{\mu \nu }^{\gamma \delta }$, etc \cite{Lobo:2008sg}. For further details on applications of certain modified gravities, we recommend the following references \cite{Nojiri:2010wj,Bamba:2012cp,Nojiri:2017ncd}. However, this class of Riemannian geometries has been recently extended to include new fundamental blocks to describe spacetime, such as the presence of the torsion scalar $T$ \cite{Aldrovandi:2013wha} and nonmetricity, $\mathbb{Q}$ \cite{Nester:1998mp}.

A fundamental result in differential geometry states that the general affine connection, which is a mathematical tool that plays a central role in defining the transport of tensors over the manifold and in defining the covariant derivative, may always be decomposed into the following three independent components \cite{Hehl:1994ue,Ortin:2015hya}:
\begin{equation}
\bar{\Gamma}{}_{\phantom{\beta}\mu\nu}^{\beta}=\Gamma_{\phantom{\beta}\mu\nu}^{\beta} +K_{\phantom{\beta}\mu\nu}^{\beta}+L_{\phantom{\beta}\mu\nu}^{\beta},
\label{conexion}
\end{equation}
where the first term corresponds to the Levi-Civita connection
\begin{equation}
\Gamma_{\phantom{\beta}\mu\nu}^{\beta}=\frac{1}{2}g^{\beta\alpha}\left(\partial_{\mu}g_{\nu\alpha}+\partial_{\nu}g_{\alpha\mu}-\partial_{\alpha}g_{\mu\nu}\right),
\label{christoffel}
\end{equation}
while the second term is known as the contorsion
\begin{equation}
K_{\phantom{\alpha}\mu\nu}^{\beta}=\frac{1}{2}T_{\phantom{\alpha}\mu\nu}^{\beta}+T_{(\mu\phantom{\beta}\nu)}^{\phantom{(\mu}\beta},\label{tns_tor} 
\end{equation}
and is defined in terms of the torsion tensor
\begin{equation}
T_{\phantom{\beta}\mu\nu}^{\beta}=2\varGamma_{\phantom{\beta}\left[\mu\nu\right]}^{\beta}=-T_{\phantom{\beta}\nu\mu}^{\beta}.
\end{equation}
Finally, we have the deformation tensor 
\begin{equation}
L_{\phantom{\alpha}\mu\nu}^{\beta}=\frac{1}{2}Q_{\phantom{\alpha}\mu\nu}^{\beta}-Q_{(\mu\phantom{\beta}\nu)}^{\phantom{(\mu}\beta}=L_{\phantom{\beta}\nu\mu}^{\beta},\label{disf}
\end{equation}
which is defined by the divergence of the metric tensor, i.e. the non-metricity tensor
\begin{equation}
 Q_{\beta\mu\nu}\equiv \nabla_{\beta}g_{\mu\nu}\neq 0,   \label{tns_nmetric}
\end{equation}
where~$\nabla_\beta$ denotes the covariant derivative, which is given with respect to the affine connection~\eqref{conexion}. 

In fact, by imposing assumptions on the affine connection, one specifies the metric-affine geometry.  Thus,  in the standard formulation of General Relativity (GR), one assumes a Levi-Civita connection, which implies a vanishing torsion and nonmetricity, while in the Teleparallel Equivalent of GR (TEGR), based on the torsion scalar $T$, one uses the Weitzenb\"{o}ck connection, implying zero curvature and nonmetricity \cite{Aldrovandi:2013wha,Cai:2015emx}. Another equivalent formulation of GR is  the symmetric teleparallel equivalent of GR (STEGR), where it is the nonmetricity tensor $\mathbb{Q}$ that describes the gravitational interaction \cite{Nester:1998mp,BeltranJimenez:2017tkd}. These three frameworks have been denoted the ``Geometric Trinity of Geometry'' \cite{BeltranJimenez:2019esp}.

As interesting extension of STEGR, namely, $f(\mathbb{Q})$ gravity, is related with the inclusion of a boundary term $\mathbb{B}_Q$. It was found that the resulting theory, $f(\mathbb{Q},\mathbb{B}_Q)$ gravity, is dynamically equivalent to $f(R)$ gravity,  with $ f(\mathbb{Q},\mathbb{B}_Q)=f(\mathbb{Q}-\mathbb{B}_Q)$ \cite{Capozziello:2023vne}. In fact, it was also shown that in extended teleparallel gravity $f(T)$, gravity, one also has $f(T,\tilde{B})=f(-T+\tilde{B})$ gravity  (where $\mathbb{B}_Q\neq \tilde{B}$) \cite{Bahamonde:2015zma}.  Thus, it was suggested  \cite{Capozziello:2023vne} that in this perspective, considering boundary terms in $ f(\mathbb{Q})$ gravity represents the last ingredient towards an ``Extended Geometric Trinity of Gravity'', where $f(R)$, $f(T,\tilde{B})$, and $f(\mathbb{Q},\mathbb{B}_Q)$ can be dealt with under the same standard. 

Furthermore, there are several relevant studies on the theory $f(\mathbb{Q})$ with cosmological applications that have been investigated in \cite{Barros:2020bgg,Frusciante:2021sio,Ayuso:2020dcu,Anagnostopoulos:2021ydo,Atayde:2021pgb}. In addition to the studies mentioned earlier on this theory, we also recommend consulting the following references to deepen your understanding \cite{Harko:2018gxr,Xu:2019sbp,Solanki:2021qni,Banerjee:2021mqk,Mustafa:2021ykn,Capozziello:2022zzh,Paliathanasis:2023pqp,Dimakis:2023uib}. A study on FLRW cosmology within the scope of the theory $f(\mathbb{Q},\mathbb{B}_Q)$, considering several families  of connections, was developed  in Ref. \cite{Paliathanasis:2023kqs}. The behavior of cosmological models of dark energy described by the theory $f(\mathbb{Q},\mathbb{B}_Q)$ was investigated with perfect fluid in Ref. \cite{Pradhan:2023oqo} and with quintessence in Ref. \cite{Maurya:2024giu}.

It this work, we explore static and spherically symmetric black hole and regular black hole solutions in $f(\mathbb{Q},\mathbb{B}_Q)$ gravity coupled to nonlinear electrodynamics (NLED).  Since the discovery of non-singular black hole solutions \cite{Bardeen, Ayon-Beato:2000mjt},  a plethora of solutions have been analysed in this class of solutions (rather than provide an exhaustive list of the solutions, we refer the reader to \cite{Bronnikov:2017sgg,Bronnikov:2000vy,Junior:2015dga,Junior:2015fya,Ayon-Beato:2000mjt,Nashed:2018cth,Nashed:2020kdb,Nashed:2021pah} and the references therein).

This paper is organised in the following manner: In Sec. \ref{sec2}, we present the fundamentals of theories of gravity with non-metricity. In Sec. \ref{sec3}, we consider $f(\mathbb{Q},\mathbb{B}_Q)$ gravity coupled to NLED and present the equations of motion of $f(\mathbb{Q},\mathbb{B}_Q)$ that will be used throughout this work.  In Sec. \ref{sec4}, we find several static and spherically symmetric black hole solutions. In Sec. \ref{sec5},  we explore regular black holes, and generalize the regular Bardeen and Culetu solutions. Finally, in Sec.\ref{sec:conclusion}, we summarize our results and conclude.

\section{Theory of gravity with non-metricity}\label{sec2}

\subsection{Fundamentals}

In this section, we present the fundamentals of $f(\mathbb{Q},\mathbb{B}_Q)$ gravity, where the non-metricity tensor is given by $Q_{\beta\mu\nu}\equiv \nabla_{\beta}g_{\mu\nu}\neq 0$, as mentioned above.
The two characteristics of the non-metricity tensor are defined by their contractions:
\begin{subequations}
    \begin{align}
Q_{\alpha}=g^{\mu\nu}Q_{\alpha\mu\nu}=Q_{\alpha\phantom{\nu}\nu}^{\phantom{\alpha}\nu},	\\
\tilde{Q}_{\alpha}=g^{\mu\nu}Q_{\mu\alpha\nu}=Q_{\phantom{\nu}\alpha\nu}^{\nu}	.
    \end{align}
\end{subequations}
It is useful to define a superpotential, given by
\begin{equation}
P_{\phantom{\alpha}\mu\nu}^{\beta}=-\frac{1}{2}L_{\phantom{\alpha}\mu\nu}^{\beta}-\frac{1}{4}\left[\left(\tilde{Q}^{\beta}-Q^{\beta}\right)g_{\mu\nu}+\delta_{(\mu}^{\beta}Q_{\nu)}\right],\label{superpot}
\end{equation}
so that the nonmetricity tensor~\eqref{tns_nmetric} with the superpotential~\eqref{superpot}, defines the nonmetricity scalar from the contraction as follows
\begin{equation}
    \mathbb{Q}=-Q_{\beta\mu\nu}P_{\phantom{\alpha}}^{\beta\mu\nu}.\label{scalarQ}
\end{equation}

Applying the corresponding geometric configurations where the curvature is zero, i.e. the teleparallel condition, and where the geometry has no torsion, we obtain the following relation
\begin{equation}
\overset{C}{R}{}_{\phantom{\beta}\alpha\mu\nu}^{\beta}+\overset{C}\nabla_{\mu}L_{\phantom{\beta}\nu\alpha}^{\beta}-\overset{C}\nabla_{\nu}L_{\phantom{\beta}\mu\alpha}^{\beta}+L_{\phantom{\beta}\mu\rho}^{\beta}L_{\phantom{\rho}\nu\alpha}^{\rho}-L_{\phantom{\beta}\nu\rho}^{\beta}L_{\phantom{\rho}\mu\alpha}^{\rho}=0,\label{Tns_Riem_Trans} 
\end{equation}
where $\overset{C}\nabla$ denotes the derivative that corresponds to the Christoffel symbol~\eqref{christoffel}, as well as the curvature tensor  $\overset{C}{R}{}_{\phantom{\beta}\alpha\mu\nu}^{\beta}$, which has its explicit form given by
\begin{equation}
\overset{C}{R}{}_{\phantom{\beta}\alpha\mu\nu}^{\beta}=\partial_{\alpha}\Gamma_{\phantom{\beta}\nu\mu}^{\beta}-\partial_{\nu}\Gamma_{\phantom{\beta}\alpha\mu}^{\beta}+\Gamma_{\phantom{\beta}\alpha\rho}^{\beta}\Gamma_{\phantom{\rho}\nu\mu}^{\rho}-\Gamma_{\phantom{\beta}\nu\rho}^{\beta}\Gamma_{\phantom{\rho}\alpha\mu}^{\rho}.    \label{tns_Riem}
\end{equation}
The contraction of this tensor produces the Ricci tensor
\begin{equation}   \overset{C}{R}_{\mu\nu}=\overset{C}{R}{}_{\phantom{\beta}\mu\beta\nu}^{\beta},\label{Ricci}
\end{equation}
and Eq.~\eqref{Tns_Riem_Trans} reduces to the following expression after the appropriate contractions
\begin{equation}
\overset{C}{R}=\mathbb{Q}-\overset{C}\nabla_{\beta}\left(Q_{\phantom{\beta}}^{\beta}-\tilde{Q}{}_{\phantom{\beta}}^{\beta}\right),
\label{scalar_Ric2}
\end{equation}
where 
\begin{equation}
  \overset{C}{R}= g^{\mu\nu}\overset{C}{R}_{\mu\nu},
  \label{scal_Curv}
\end{equation}
is the contraction of Eq.~\eqref{Ricci}, expressed in terms of~\eqref{christoffel}.

Thus, taking into Eq.~\eqref{scalar_Ric2} we verify that the non-metricity scalar differs from the Ricci scalar by a term of the total derivation
\begin{equation}
\mathbb{B}_Q=\overset{C}\nabla_{\beta}\left(Q^{\beta}-\tilde{Q}^{\beta}\right),
\label{boundary}
\end{equation}
which will be a fundamental quantity used throughout this work.

\subsection{Coincident Gauge}

Let us now discuss an interesting symmetry property used in STEGR that simplifies this formulation somewhat, where the coordinate transformation considered here is known as the \emph{``coincident gauge"} \cite{BeltranJimenez:2017tkd}. More specifically, a variety of flat geometry, where we have a flat connection with $T^\beta_{\phantom{\beta}\mu\nu}=0$, can be described by
\begin{equation}
\bar{\Gamma}{}_{\phantom{\beta}\mu\nu}^{\beta}=\left(\Lambda^{-1}\right)_{\phantom{\alpha}\alpha}^{\beta}\partial_{\mu}\Lambda_{\phantom{\beta}\nu}^{\alpha}.\label{con_inertial}
\end{equation}
where $\Lambda_{\phantom{\alpha}\mu}^{\alpha}$ belongs to ${\rm GL}(4, {\cal \mathbb{R}})$.

The constraint in which the torsion is zero implies \begin{equation}
\left(\Lambda^{-1}\right)_{\phantom{\alpha}\beta}^{\alpha}\partial_{[\mu}\Lambda_{\phantom{\alpha}\nu]}^{\beta}=0\,,
\end{equation}
so that it is possible to rewrite the connection~\eqref{con_inertial} from a parametrization according to the transformation $\Lambda_{\phantom{\alpha}\mu}^{\alpha}=\partial_{\mu}\xi^{\alpha}$. Consequently, we obtain
\begin{equation}
\bar{\Gamma}{}_{\phantom{\alpha}\mu\nu}^{\beta}=\frac{\partial x^{\beta}}{\partial\xi^{\rho}}\partial_{\mu}\partial_{\nu}\xi^{\rho}, \label{coinc}
\end{equation}
where $\xi^{\lambda}$ is a set of arbitrary functions of the coordinates $x^{\lambda}$. 

As mentioned above, this result makes it possible, by a suitable choice of coordinates, to make the connection flat, torsion-free, and therefore consequently vanishes. The choice of coordinates that cancels the connection is given by $x^\lambda=\xi^\lambda$, thus
\begin{equation}
    \bar{\Gamma}{}_{\phantom{\beta}\mu\nu}^{\beta}=0,
\end{equation}
due $\partial_{\mu}\partial_{\nu}\xi^{\rho}=0$.

The coordinate transformation that implies a connection described by the above result is called the \emph{``coincident  gauge"}~\cite{BeltranJimenez:2017tkd, BeltranJimenez:2019tme}, so that we have
\begin{equation}
    \Gamma_{\phantom{\beta}\mu\nu}^{\beta}=-L_{\phantom{\beta}\mu\nu}^{\beta}.
\end{equation}
The non-metricity tensor defined according to Eq.~\eqref{tns_nmetric} now becomes
\begin{equation}
 Q_{\beta\mu\nu}\equiv \partial_{\beta}g_{\mu\nu}. \label{tns_nmetric2}
\end{equation}

The coincident gauge is an extremely useful symmetry, which will be used to develop the solutions outlined throughout this work.

\subsection{$f(\mathbb{Q})$ gravity}

The STEGR describes the gravitational interaction by the non-metricity tensor,~$Q_{\alpha\mu\nu}$, and the action is defined as:
\begin{equation}
S_{\text{STEGR}}=\int\sqrt{-g}d^{4}x\left(\mathbb{Q}+2\kappa^2\mathcal{L}_m\right)\label{action_Q},
\end{equation}
where $\kappa^2=8\pi G/c^4$,  $G$ is the gravitational constant and $\mathcal{L}_{m}$ is the Lagrangian of the matter field.

Note that from the relation \eqref{scalar_Ric2}, the action of STEGR differs from the Einstein-Hilbert action of GR by a boundary term $\mathbb{B}_Q$, given by Eq.~\eqref{boundary}, which means that STEGR is an equivalent formulation of GR.


A nonlinear generalization of STEGR is a proposal in which the action is described as follows
\begin{equation}
S_{ f}=\int\sqrt{-g}d^{4}x\left[f\left(\mathbb{{Q}}\right)+2\kappa^2\mathcal{L}_{m}\right],\label{action_f(Q)}
\end{equation}
where $f(\mathbb{Q})$ is an arbitrary function of the nonmetricity scalar $\mathbb{Q}$.

However, this theory raises some issues, for instance, if one chooses the coincident gauge with static and spherical symmetry, the equations of motion imply a theory with a linear $f(\mathbb{Q})$ function or constant scalar non-metricity, as verified in~\cite{Junior:2023qaq}. This is undoubtedly due to the use of a geometric object which is not a Lorentz invariant.
To overcome this challenge, a theory must be constructed that accounts for this invariance.

\subsection{$f(\mathbb{Q},\mathbb{B}_Q)$ gravity}

In this work, we develop solutions based on a recent proposal by extending the theory $f(\mathbb{Q})$, namely, $f(\mathbb{Q},\mathbb{B}_Q)$ gravity~\cite{Capozziello:2023vne, De:2023xua}, where $\mathbb{B}_Q$ is the boundary term given by Eq. (\ref{boundary}). In this approach, the action is defined as:
\begin{equation}
S_{ f(\mathbb{Q},\mathbb{B}_Q)}=\int\sqrt{-g}d^{4}x\Big[f\left(\mathbb{{Q}},\mathbb{B}_Q\right) +2\kappa^2\mathcal{L}_{m}\Big]\label{action_f(QB)}.
\end{equation}

The variation of the action \eqref{action_f(QB)} with respect to the metric tensor, yields the following equation of motion:
\begin{eqnarray}
 f_{Q}\left(\mathbb{Q}\right)\overset{C}{G}_{\mu\nu}-\frac{1}{2}g_{\mu\nu}\bigr[f(\mathbb{Q})-f_{Q}\left(\mathbb{Q}\right)\mathbb{Q}-f_{B}\left(\mathbb{B}_Q\right)\mathbb{B}_Q\bigl]
	 +2P_{\phantom{\alpha}\mu\nu}^{\alpha}\partial_{\alpha}\bigr[f_{Q}\left(\mathbb{Q}\right)+f_{B}\left(\mathbb{B}_Q\right)\bigl]
	\nonumber \\	 
	 -g_{\mu\nu}\overset{C}{\square} f_{B}\left(\mathbb{B}_Q\right) 
+\overset{C}{\nabla}_{\mu}\overset{C}{\nabla}_{\nu}f_{B}\left(\mathbb{B}_Q\right)=\kappa^{2}\Theta_{\mu\nu},
\label{eq_f(QB)}
\end{eqnarray}
 where $\overset{C}{G}_{\mu\nu}$ is the Einstein tensor of GR and we denote $f_{Q}\left(\mathbb{Q},\mathbb{B}_Q\right)=\partial f\left(\mathbb{Q},\mathbb{B}_Q\right)/\partial\mathbb{Q}$, 
 $f_{B}\left(\mathbb{Q},\mathbb{B}_Q\right)= \partial f\left(\mathbb{Q},\mathbb{B}_Q\right)/\partial\mathbb{B}_Q$, the energy-momentum tensor is  $\Theta_{\mu\nu}$  and $\square=\nabla^{\mu}\nabla_{\mu}$.  Note that the contribution of the boundary term leads to the existence of fourth order field equations in this theory \cite{Capozziello:2023vne}.

Now, the variation of the action \eqref{action_f(QB)} with respect to the affine component provides the following field equation
\begin{equation}
    \nabla_{\mu}\nabla_{\nu}\Bigl[\sqrt{-g}P_{\phantom{\mu\nu}\alpha}^{\mu\nu}\Bigl(f_{Q}\left(\mathbb{Q}\right)+f_{B}\left(\mathbb{B}_Q\right) \Bigr)\Bigr]=0.
\end{equation}

If we substitute $\mathbb{B}_Q=0$ into the field equation \eqref{eq_f(QB)}, leading to $f_{B}(\mathbb{Q},\mathbb{B}_Q)$=0, we arrive at the field equations of $f(\mathbb{Q})$ gravity.  
Note that the resulting equations of motion can be rewritten as \cite{DAmbrosio:2021zpm, Zhao:2021zab, Lin:2021uqa}
\begin{eqnarray}
 f_{{Q}}\left(\mathbb{Q}\right)\overset{C}{G}_{\mu\nu}-\frac{1}{2}g_{\mu\nu}\Bigl(f(\mathbb{Q})-f_{Q}\left(\mathbb{Q}\right)\mathbb{Q}\Bigr)
 +2P_{\phantom{\alpha}\mu\nu}^{\alpha}\partial_{\alpha}f_{Q}\left(\mathbb{Q}\right)=\kappa^2\Theta_{\mu\nu}.
\end{eqnarray}

An interesting property is that $f(R)$ gravity is recovered by considering $R=\mathbb{Q}-\mathbb{B}_Q$, i.e., $f(R)=f(\mathbb{Q},\mathbb{B}_Q)=f(\mathbb{Q}-\mathbb{B}_Q)$, which yields the following field equations
\begin{eqnarray}
    \left(\overset{C}{G}_{\mu\nu}+\frac{1}{2}g_{\mu\nu}R+g_{\mu\nu}\overset{C}{\square}-\overset{C}{\nabla}_{\mu}\overset{C}{\nabla}_{\nu}\right)f_{R}\left(R\right)  
    -\frac{1}{2}g_{\mu\nu}f\left(R\right)=\kappa^{2}\Theta_{\mu\nu}.\label{eq_f(Q)}
\end{eqnarray}

\section{$f(\mathbb{Q},\mathbb{B}_Q)$ gravity coupled to non-linear electrodynamics}\label{sec3}

\subsection{Action and metric}

In this section, we consider $f(\mathbb{Q},\mathbb{B}_Q)$ gravity coupled to NLED. The action is given by
\begin{eqnarray}
    S=\int d^{4}x\sqrt{-g}\Bigl[f(\mathbb{Q},\mathbb{B}_{Q}) + 2\kappa^2 {\cal L}_{\rm NLED}({F})\Bigr],   \label{action_f_NED_phi}
\end{eqnarray}
where ${\cal L} _{\rm NLED}(F)$ is the Lagrangian density describing the NLED that depends on the electromagnetic scalar $F$ as
\begin{equation}
F=\frac{1}{4}F^{\mu\nu}F_{\mu\nu},    \label{F}
\end{equation}
and the electromagnetic field $F_{\mu \nu} $, defined as 
\begin{equation}
   F_{\mu \nu} = \partial_\mu A_\nu -\partial_\nu A_\mu, 
\end{equation}
 is the Maxwell-Faraday antisymmetric tensor, where ${A_\beta}$ is the magnetic vector potential. 
 
Varying the action \eqref{action_f_NED_phi} with respect to $A_\beta$, we have
\begin{equation}
    \nabla_\mu [{\cal L}_F F^{\mu\nu}]=\partial_\mu [\sqrt{-g} {\cal L}_F F^{\mu\nu}]=0,\label{sol2}
\end{equation}
where we denote ${\cal L}_F = \partial {\cal L} _{\rm NLED}(F)/\partial F$.

Finally, varying the action \eqref{action_f_NED_phi} with respect to the metric, we obtain the equations of motion \eqref{eq_f(QB)} and the matter contributions expressed by the energy-momentum tensor, with the following contribution
\begin{equation}
    \Theta_{\phantom{\mu}\nu}^\mu = \overset{F}{\Theta}   
 {_{\phantom{\mu}\nu}^\mu}
 \label{TEM},
\end{equation}
where the nonlinear electromagnetic energy-momentum tensor $\overset{F}{\Theta}{}_{\phantom{\mu}\nu}^{\mu}$ is defined as
\begin{equation}
 \overset{F}{\Theta}{}_{\phantom{\mu}\nu}^{\mu}= \delta^\mu_\nu{\cal L}_{\rm NLED} (F)  +{\cal L}_F F^{\mu\alpha} F_{\nu\alpha}. \label{TEMF}
\end{equation}

The solutions that we discuss later in the context of black holes and regular black holes will be obtained using the following static and spherically symmetric metric
\begin{equation}
ds^2=A(r)dt^2-\frac{1}{A(r)}dr^2-\Sigma^2(r)\left(d\theta^{2}+\sin^{2}\theta \, d\phi^{2}\right),\label{metric}
\end{equation}
where $A(r)$ and $\Sigma(r)$ are functions of the radial coordinate $r$.

Here, we only consider the magnetic charge, so that the only non-zero component of the tensor $F_{\mu\nu}$ is
\begin{equation}
    F_{23}=q \sin\theta,
\end{equation}
and the electromagnetic scalar is given by
\begin{equation}
    F=\frac{q^2}{2 \Sigma^4(r)}.
\end{equation}

In this way, using the metric \eqref{metric}, we can describe boundary term \eqref{boundary} as follows
\begin{eqnarray}
\mathbb{B}_Q(r)=A''(r)+
\frac{6 A'(r) \Sigma '(r)+4 A(r) \Sigma ''(r)}{\Sigma (r)} 
+\frac{4A(r)\Sigma'^{2} (r)-2}{\Sigma(r)} ,\label{B}  
\end{eqnarray}
and the nonmetricity scalar \eqref{scalarQ} now becomes
\begin{equation}
\mathbb{Q} (r)=-\frac{2 \Sigma '(r) \Bigl(\Sigma (r) A'(r)+A(r) \Sigma '(r)\Bigr)}{\Sigma ^2 (r)}.\label{Q}
\end{equation}

\subsection{Equations of motion}

Thus, taking into account the field equation \eqref{eq_f(QB)}, the line element \eqref{metric}, and the energy-momentum tensor \eqref{TEM}, the equations of motion for $f(\mathbb{Q},\mathbb{B}_{Q})$ gravity coupled with NLED are given by
\begin{eqnarray}
 \frac{1}{2}f_{Q}(r)\left[\mathbb{Q}(r)-\frac{2\Bigl(\Sigma(r)A'(r)\Sigma'(r)+A(r)\Sigma'(r)^{2}-1\Bigr)}{\Sigma(r)^{2}}\right]
 + \frac{1}{2}\mathbb{B}_Q(r)f_{B}(r)
 	\nonumber\\	
 	 -\frac{f(r)}{2}
-\frac{f_{B}^{\prime}(r)\Big(\Sigma(r)A'(r)+4A(r)\Sigma'(r)\Big)}{2\Sigma(r)}
 =	\kappa^{2}{\cal L}_{\rm NLED}(r),
\end{eqnarray}
\begin{eqnarray}
 \frac{1}{2} A(r) \cot \theta  \left(f_B'(r)+f_Q'(r)\right)=0,\label{eq12}
\end{eqnarray}
\begin{eqnarray}
\frac{\cot \theta  \left(f_B'(r)+f_Q'(r)\right)}{2 \Sigma^2 (r)}=0, \label{eq21}
\end{eqnarray}
%
%
\begin{eqnarray}
  \frac{1}{2}f_{Q}(r)\left\{\mathbb{Q}(r)-\frac{2\left[\Sigma(r)\Bigl(A^{\prime}(r)\Sigma^{\prime}(r)+2A(r)\Sigma^{\prime\prime}(r)\Bigr)+A(r)\Sigma^{\prime2}(r)-1\right]}{\Sigma(r)^{2}}\right\}-\frac{2A(r)\Sigma^{\prime}(r)\left(f_{B}^{\prime}(r)+f_{Q}^{\prime}(r)\right)}{\Sigma(r)}
  	\nonumber\\ \label{eqm}
  -\frac{f(r)}{2}	
+\frac{1}{2}\left[-\frac{f_{B}^{\prime}(r)\Bigl(\Sigma(r)A'(r)+4A(r)\Sigma^{\prime}(r)\Bigr)}{\Sigma(r)}-2A(r)f_{B}^{\prime\prime}(r)+\mathbb{B}_Q(r)f_{B}(r)\right]	=	\kappa^{2}{\cal L}_{\rm NLED}(r),
\end{eqnarray}
\begin{eqnarray}
&& -\frac{f_{B}^{\prime}(r)\left(\Sigma(r)A^{\prime}(r)+A(r)\Sigma^{\prime}(r)\right)}{\Sigma(r)}-A(r)f_{B}^{\prime\prime}(r)+\frac{1}{2}f_{Q}(r)\left(\mathbb{Q}(r)-\frac{\Sigma(r)A^{\prime\prime}(r)+2A^{\prime}(r)\Sigma^{\prime}(r)+2A(r)\Sigma^{\prime\prime}(r)}{\Sigma(r)}\right)
	\nonumber\\	
&& 
\hspace{0.5cm} +\frac{1}{2}\mathbb{B}_Q(r)f_{B}(r)
 -\frac{1}{2}\left[f(r)+\frac{\Big(\Sigma(r)A'(r)+2A(r)\Sigma'(r)\Big)\Big(f_{B}^{\prime}(r)+f_{Q}^{\prime}(r)\Big)}{\Sigma(r)}\right]	
 =\kappa^{2}\left({\cal L}_{\rm NLED}(r)-\frac{q^{2}L_{F}(r)}{\Sigma^{4}(r)}\right),\label{eqm2}
\end{eqnarray}
where the prime denotes a derivative with respect to the radial coordinate $r$.

\section{Black hole solutions in $f(\mathbb{Q,B_Q})$ gravity}
\label{sec4}

In this section, in order to deduce black hole solutions, consider the following metric
\begin{equation}
ds^2=A(r)dt^2-\frac{1}{A(r)}dr^2-r^2\left(d\theta^{2}+\sin^{2}\theta\,d\phi^{2}\right),\label{m_bn}
\end{equation}
where we have used $ \Sigma(r) = r$ for simplicity.

Thus the boundary term \eqref{B} is now given as 
\begin{equation}
    \mathbb{B} _Q(r)=-\frac{r^2 A''(r)+6 r A'(r)+4 A(r)-2}{r^2},
    \label{B_BN}
\end{equation}
and the non-metricity scalar \eqref{Q} takes the form
\begin{equation}
    \mathbb{Q}(r)=-\frac{2  \Bigl(r A'(r)+A(r) \Bigr)}{r^2},
    \label{Q_BN}
\end{equation}
where Eqs. \eqref{B_BN} and \eqref{Q_BN} satisfy  condition \eqref{scalar_Ric2}.

To obtain black hole solutions, and in order to ensure that our solutions preserve Lorentz symmetry, we assume the following relation, throughout this work: 
\begin{equation}
    f_Q = -f_B.\label{Symmetry2}
\end{equation}
Equation \eqref{Symmetry2} contains the particular case of the symmetric teleparallel theory $f(\mathbb{Q},\mathbb{B}_Q)=\mathbb{Q}-\mathbb{B}_Q$, and even if $f_Q=1$ and $f_B=-1$ this equation is identically satisfied. 
In the coincident gauge, condition \eqref{Symmetry2} must always be enforced, where the equations of motion \eqref{eq12} and \eqref{eq21} are identically satisfied. If Eq. \eqref{Symmetry2} is not imposed for the entire radial coordinate range, two of the equations of motion, i.e., \eqref{eq12} and \eqref{eq21}, will not be satisfied for some value or interval of $r$. We conclude that condition \eqref{Symmetry2} is a necessary imposition for the coincident gauge in the theory $f(\mathbb{Q},\mathbb{B}_Q)$ with static spherically  symmetry. It is likely that this condition can be relaxed for metrics that also depend on time, such as the Lemaître-Tolman-Bond metric, as is the case with the $f(T)$ theory 
 \cite{Rodrigues:2012wt}.

 Thus,  solving the equations of motion (\ref{eqm})-(\ref{eqm2}), we obtain the following quantities:
\begin{eqnarray}
 {\cal L} _{\rm NLED} (r)= \frac{1}{2 \kappa ^2}\left[ \frac{r^2 f'(r) \Bigl(r A''(r)+2 A'(r)\Bigr)}{r^3 A^{3}(r)+6 r^2 A''(r)-2 r A'(r)-8 A(r)+4}
-f(r) \right], \label{L2_BN}
\end{eqnarray}
\begin{eqnarray}
 {\cal L} _F (r)= \frac{r^5 \Bigl(r^2 A''(r)-2 A(r)+2\Bigr) f'(r)}{2 \kappa ^2 q^2 \Bigl(r^3 A^{3}(r)+6 r^2 A''(r)-2 r A'(r)-8 A(r)+4\Bigr)}.\label{LF2_BN}
\end{eqnarray}


\subsection{First model of black hole}\label{BN1}

It is an extremely difficult task to solve the equations of motion, and therefore we propose several simplifications. 
As a starting point, we model the boundary term described by Eq.~\eqref{B_BN} and assume that it is a regular quantity, so that we can solve the differential equation to find an expression for the metric function $A(r)$. 

Thus, consider the following boundary term
\begin{equation}
    \mathbb{B}_Q(r)=\frac{b_0}{r^2},\label{B_BN1}
\end{equation}
so that the black hole solution can therefore be determined from Eq.~\eqref{B_BN} as follows:
\begin{equation}
    -\frac{r^2 A''(r)+6 r A'(r)+4 A(r)-2}{r^2}=\frac{b_0}{r^2}.
\end{equation}
which provides the following metric function
\begin{equation}
    A(r)=\frac{2-b_0}{4}+\frac{b_1}{r^4}-\frac{2 M}{r},\label{A1_BN}
\end{equation}
where $b_0$ an $b_1$ are constants. 
Note that we obtain the Schwarzschild metric function by imposing $b_0=-2$ and $b_1=0$ into Eq.~\eqref{A1_BN}.  In the limit $r \rightarrow 0$, the metric function diverges, while in the limit $r \rightarrow \infty$, it is asymptotically flat. If we set $b_0=-2$, the  metric function \eqref{A1_BN} is asymptotically Minkowskian.

The non-metricity scalar for this solution is given by
\begin{equation}
    \mathbb{Q}(r)=\frac{b_0-2}{2 r^2}+\frac{6 b_1}{r^6}.\label{Q_BN1}
\end{equation}
Note that the non-metricity scalar and the boundary term satisfy the condition~\eqref{scalar_Ric2}.
Although $\mathbb{Q}(r)$ and $\mathbb{B}_Q(r)$ are divergent in the limit $r\to0$, the combination $\mathbb{Q}(r)-\mathbb{B}_Q(r)=R$  is what characterizes whether the solution is regular or not. Thus, as $\mathbb{Q}(r)-\mathbb{B}_Q(r)$ is singular at $r=0$, the geometry of the spacetime is that of a singular solution (we refer the reader to the comment in Ref.~\cite{Junior:2023qaq} under Eq. (46)).


To determine the existence of event horizons in a given solution, it is necessary to solve the following equation:
\begin{equation}
     A(r_H)=0,\label{rH}
\end{equation}
where $r_{H}$ is the radius of the horizon. 

In addition, we can use the following condition to obtain the values of the critical parameters
\begin{equation}
    \frac{d A{(r)}}{dr}\bigg|_{r=r_H}=0.\label{der_a}
\end{equation}
We can therefore solve equations \eqref{rH} and \eqref{der_a} simultaneously to obtain algebraic solutions for the radius of the event horizon and the value of a certain critical parameter in the model. However, depending on the model, it may be difficult to find analytical solutions for these quantities. Therefore, we will approach this problem numerically by assigning values to the constants to investigate the presence of event horizons through graphical representations.

\begin{figure}[htbp]
\centering
\includegraphics[scale=0.4]{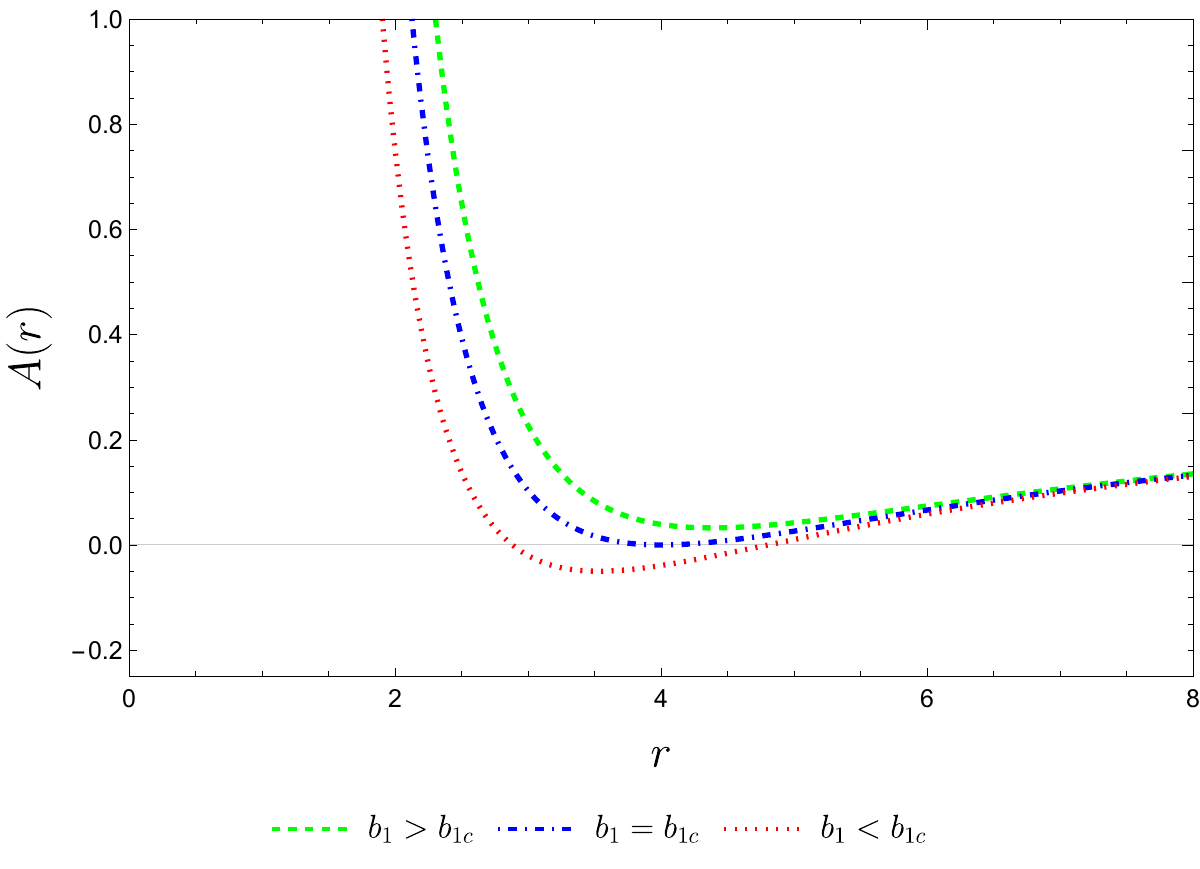}
\caption{The plot depicts the metric function \eqref{A1_BN}, i.e., $A(r)$, for the specific values of $b_0=0.5$ and $M =1.0$, where Eq. \eqref{der_a} determines the critical value of the constant $b_1$, given by $b_{1c} = 32.0$.  For $b_1 > b_{1c}$ there is no horizon, for $b_{1} = b_{1c}$, there is only one horizon and for $b_{1} < b_{1c}$, there are two horizons.} 
\label{Axr_BN}
\end{figure}

Therefore, to find numerical solutions, we set the following values for the parameters: $b_0=0.5$ and $M=1.0$. If the two conditions \eqref{rH} and \eqref{der_a} are satisfied simultaneously, we obtain the critical parameter value $b_{1c}=32$. Figure \ref{Axr_BN} shows the behavior of the metric function~\eqref{A1_BN}, where $b_{1c}$ takes the values $b_1>b_{1c}$, $b_1=b_{1c}$ and $b_1<b_{1c}$. For $b_1>b_{1c}$ there are no event horizons; for $b_1=b_{1c}$ there is only one horizon; and for $b_1<b_{1c}$ two event horizons are present.  We can see in Fig. \ref{Axr_BN} that all the curves for the different values of $b_1$ with $r\gg1$ match.


The quantities ${\cal L}_{\rm NLED}$ and ${\cal L}_F$ take the form
\begin{equation}
{\cal L}_{\rm NLED}(r)=-\frac{1}{2 \kappa ^2} \Biggl(   f(r)-\frac{6 b_1 f'(r)}{b_0 r^3}\Biggr),\label{L_BN1}
\end{equation}
\begin{equation}
{\cal L}_F(r)=\frac{r \Bigl[\bigl(b_0+2\bigr) r^4+36 b_1\Bigr] f'(r)}{8 b_0 \kappa ^2 q^2},\label{LF_BN1}
\end{equation}
respectively, and indeed, these obey the consistency relation
\begin{equation}
    {\cal L}_F-\frac{\partial {\cal L} _{\rm NLED}}{\partial r} \bigg(\frac{\partial F}{\partial r}\bigg)^{-1}=0.\label{RC}
\end{equation}

If we solve the consistency condition~\eqref{RC}, we get the following solution
\begin{equation}
    f(r)=-\frac{f_0 r \sqrt[4]{3}\; \Gamma\!\left(\frac{1}{4},-\frac{(b_0-2) r^4}{48 b_1}\right)}{2 \sqrt[4]{-\frac{(b_0-2) r^4}{b_1}}}+f_1,
\end{equation}
where $f_0$ and $f_1$ are the constants of the integration.

Consequently, we can express the function as follows
\begin{equation}
   f(\mathbb{Q},\mathbb{B}_Q) =-\frac{\sqrt{\frac{\xi}{\mathbb{Q}-\mathbb{B}_{Q}}}}{4\sqrt{6}}\;
   \text{E}_{\frac{3}{4}}\left(-\frac{(b_{0}-2)\xi^{2}}{1728\,b_{1}\left(\mathbb{Q}-\mathbb{B}_{Q}\right)^{2}}\right).\label{fQB}
\end{equation}
where ${\textrm E}_k$ is the exponential integral function 
\begin{equation}
{\textrm E}_{k}\left(w\right)=\int_{1}^{\infty}\frac{e^{-wt}}{t^{k}}dt,
\end{equation}
and $\xi$ is given by
\begin{eqnarray}
\xi &=& \sqrt[3]{36\sqrt{b_{0}(\mathbb{Q}-\mathbb{B}_Q)^{2}\left(324b_{1}(\mathbb{Q}-\mathbb{B}_Q)^{2}-(b_{0}+2)^{3}\right)}-(b_{0}+2)^{3}+648b_{1}(\mathbb{Q}-\mathbb{B}_Q)^{2}}	
	\nonumber\\
&&+\frac{(b_{0}+2)^{2}}{\sqrt[3]{36\sqrt{b_1(\mathbb{Q}-\mathbb{B}_Q)^{2}\left(324 b_1(\mathbb{Q}-\mathbb{B}_Q)^{2}-(b_{0}+2)^{3}\right)}-(b_{0}+2)^{3}+648b_{1}(\mathbb{Q}-\mathbb{B}_Q)^{2}}}-b_{0}-2.
\end{eqnarray}

By choosing $b_0=-2$ and $b_1=1$ in Eq. \eqref{fQB}, we get the form of 
\begin{equation}
    f(\mathbb{Q},\mathbb{B}_Q) = -\frac{\sqrt[6]{3}\; {\textrm E}_{\frac{3}{4}}\left(\frac{1}{2 \sqrt[3]{6} (\mathbb{Q}-\mathbb{B}_Q)^{2/3}}\right)}{2\ 2^{5/6} \sqrt[6]{\mathbb{Q}-\mathbb{B}_Q}}\label{fQB2}
\end{equation}
that permits us see more clearly that the Eq.~\eqref{fQB2} satisfies $f_Q=-f_B$.

\begin{figure}[h]
\centering
\includegraphics[scale=0.4]{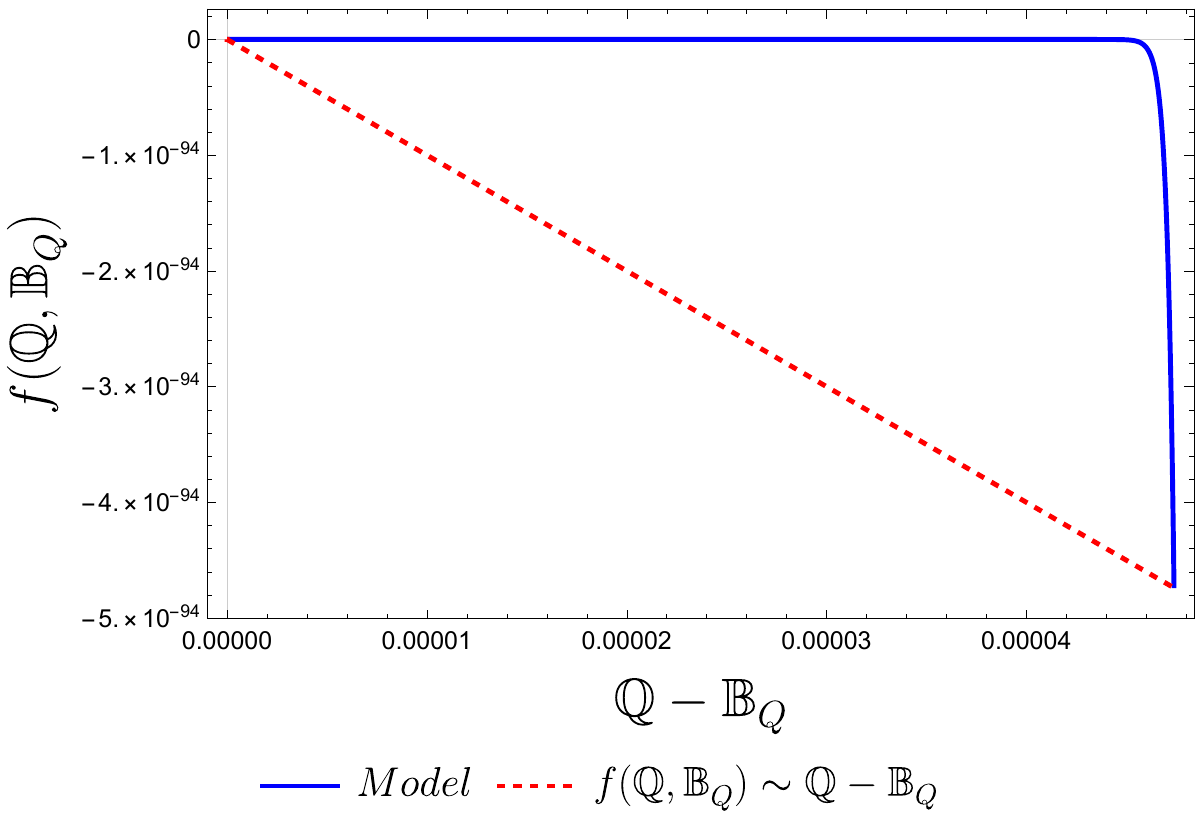}
\caption{Graphic representation of function $f(\mathbb{Q},\mathbb{B}_Q)$, described by expression~\eqref{fQB}. Where we consider $M=10$, $b_0=-2$ and $b_1=1$.  The blue curve represents the behavior of our model for $(\mathbb{Q}-\mathbb{B}_Q)\ll1$, while the dashed red line represents the linear case.} 
\label{f(QB)_BN}
\end{figure}

We have investigated the asymptotic behavior of function \eqref{fQB} for small and very large values of $\mathbb{Q}-\mathbb{B}_Q$. We found that there is no linear behavior in these two cases. To illustrate this, in Fig. \ref{f(QB)_BN} we compared function \eqref{fQB} for small values of $\mathbb{Q}-\mathbb{B}_Q$ using the blue curve with the linear case represented by a dashed red line. However, if we analyze very small intervals of $\mathbb{Q}-\mathbb{B}_Q$, the behavior we observe is practically linear.

From the function $f(r)$ at our disposal, the quantities are now expressed as:
\begin{eqnarray}
    && f_B(r)=-\frac{r^3 e^{\frac{(b_0-2) r^4}{48 b_1}}}{2 b_0},\\
    && {\cal L}_{\rm NLED}(r)=\frac{r \, {\textrm E}_{\frac{3}{4}}\left(-\frac{(b_0-2) r^4}{48 b_1}\right)}{8 \kappa ^2}+\frac{3 b_1 e^{\frac{(b_0-2) r^4}{48 b_1}}}{b_0 \kappa ^2 r^3},\label{L2_BN1}\\
    && {\cal L}_F(r)=\frac{r e^{\frac{(b_0-2) r^4}{48 b_1}} \left((b_0+2) r^4+36 b_1\right)}{8 b_0 \kappa ^2 q^2},
    \label{LF2_BN1}
\end{eqnarray}
where Eqs.~\eqref{L2_BN1} and~\eqref{LF2_BN1} satisfy the consistency condition~\eqref{RC}.

\begin{figure}[h]
\centering
\includegraphics[scale=0.4]{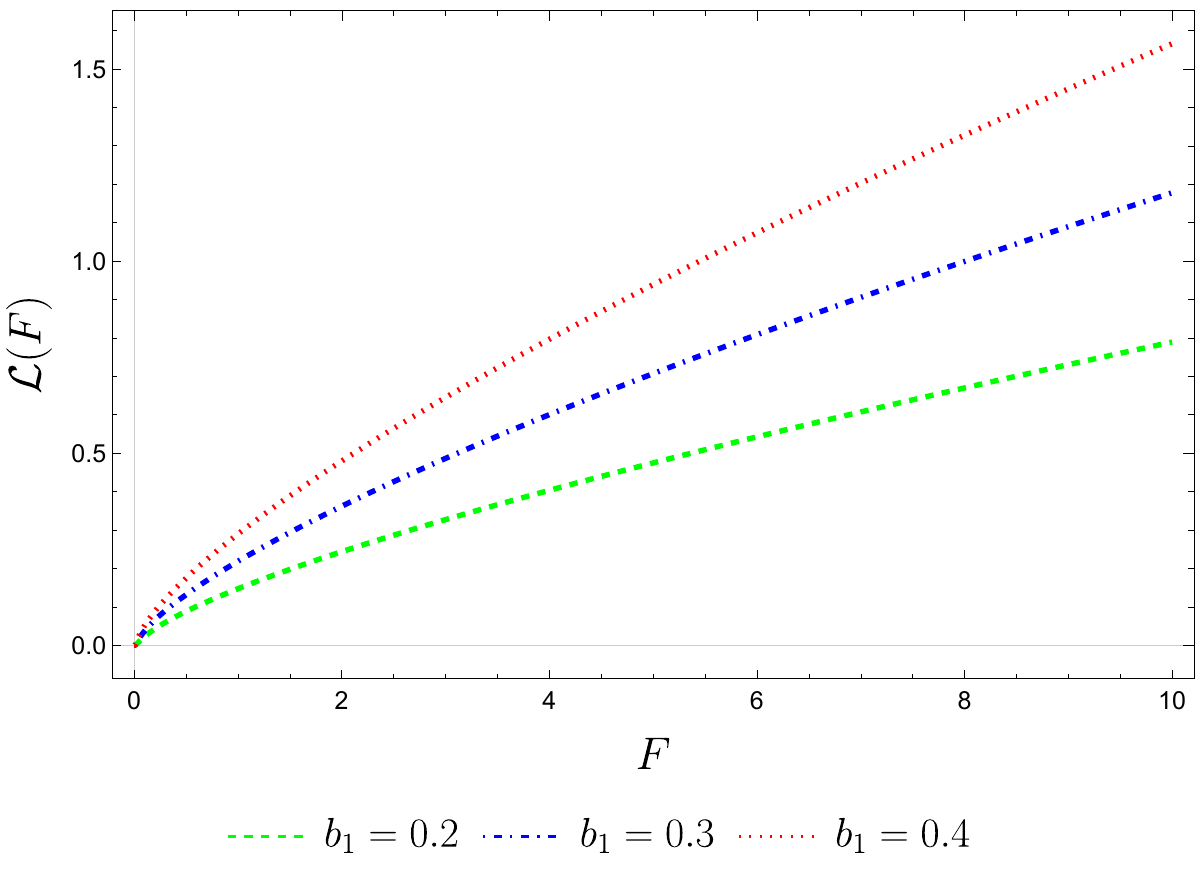}
\caption{Graphic representation of ${\cal L}_{\rm NLED}(F)$, described by expression~\eqref{LF3_BN1}, with respect to $F$. Where we consider $q=0.7$ and $b_0=0.5$.} 
\label{LxF_BN}
\end{figure}

Finally, using Eq.~\eqref{F}, we derive the expression for $r(F)$, and consequently, we are able to express the electromagnetic Lagrangian in terms of $F$, which is given by
\begin{equation}
    {\cal L} _{\rm NLED} (F) =\frac{b_0 q^2 \, {\textrm E}_{\frac{3}{4}}\left(-\frac{(b_0-2) q^2}{96 b_1 F}\right)+48 b_1 F e^{\frac{(b_0-2) q^2}{96 b_1 F}}}{8 \sqrt[4]{2} b_0 \sqrt[4]{F} \kappa ^2 q^{3/2}}.\label{LF3_BN1}
\end{equation}
We illustrate the behavior of the Lagrangian~\eqref{LF3_BN1} in Fig.~\ref{LxF_BN} for $q=0.7$, $b_0=0.5$ and three different values of $b_1$.
A note that setting $b = 2$ is an inconsistent value, as it lies outside the domain of our solution.

Let us now analyze the asymptotic limits of the Lagrangian described by Eq. \eqref{LF3_BN1}. For instance, for $F\gg1$, we observe that
\begin{equation}
    {\cal L} _{\rm NLED} (F)\sim \frac{3\ 2^{3/4} b_1 F^{3/4}}{b_0\kappa ^2 q^{3/2}}.\label{LF4_BN1}
\end{equation}
Whereas for $F\ll 1$,
\begin{equation}
 {\cal L} _{\rm NLED} (F)\sim 
\exp\left[\frac{\left(b_{0}-2\right)q^{2}}{96b_{1}F}\right]\left[-\frac{3\left(2^{3/4}\left(b_{0}+2\right)b_{1}\right)F^{3/4}}{\left(b_{0}-2\right)b_{0}q^{3/2}\kappa^{2}}\right].\label{LF5_BN1}
\end{equation}
Therefore, based on results \eqref{LF4_BN1} and \eqref{LF5_BN1}, we find that we have not recovered the linear case.

If we analyze the graph in Fig.~\ref{LxF_BN}, we find that the Lagrangian ${\cal L}_{\rm NLED} (F)$ described by Eq.~\eqref{LF3_BN1}, is nonlinear. Another important feature is that it is not ``function multivalued". We also note that this Lagrangian is a bijective function, since each element in the domain maps to only one element in the codomain. This is a peculiarity of magnetically charged solutions where we have a unique representation, which means that a single Lagrangian is sufficient to describe a single solution. This aspect does not apply to electrically charged solutions, where the Lagrangians are no longer bijective, as for one value of $F$ there can be one or more values of ${\cal L}_{\rm NLED}(F)$ (see Ref. \cite{Rodrigues:2020pem,Bronnikov:2000vy}). Furthermore, we see in the graph that the Lagrangian ${\cal L}_{\rm NLED}(F)$ has no peaks, which is another characteristic feature of magnetically charged solutions. We can also observe that by increasing the value of $b_1$
 while maintaining $q$ and $b_0$ fixed, the Lagrangian ${\cal L}_{\rm NLED}(F)$ assumes increasing values.

\subsection{Second model of black hole}\label{BN2}

In this model we adopt the symmetry presented in Eq. \eqref{Symmetry2}, as before, since it gives us analytical solutions.  We model the non-metricity scalar from Eq.~\eqref{Q_BN} so that it becomes a regular quantity, which allows us to obtain a differential equation to find an expression for $A(r)$. 
For instance, consider the following non-metricity scalar
\begin{equation}
    \mathbb{Q}(r)=\frac{q_0}{r^2+r_0^2}\,,
    \label{Q2_BN}
\end{equation}
so that the solution is therefore characterized by Eq.~\eqref{Q_BN},  and is given by 
\begin{equation}
    -\frac{2 \Bigl(r A'(r)+A(r)\Bigr)}{r^2}=\frac{q_0}{r^2+r_0^2}.
\end{equation}
which yields the following metric function
\begin{equation}
    A(r)=-\frac{2 M}{r}-\frac{q_0}{2} \left[1-\frac{r_0}{r}  \tan ^{-1}\left(\frac{r}{r_0}\right)\right].
    \label{A2_BN}
\end{equation}
where $q_0$ an $r_0$ are constants. The metric function~\eqref{A2_BN} diverges when we take the limit $r\to0$, however, for the limit $r\rightarrow\infty$, this $A(r)$ function is asymptotically flat. If we set $q_0=-2$ and $r_0=0$ in Eq. \eqref{A2_BN}, we recover the Schwarzschild solution.

\begin{figure}[htbp]
\centering
\includegraphics[scale=0.4]{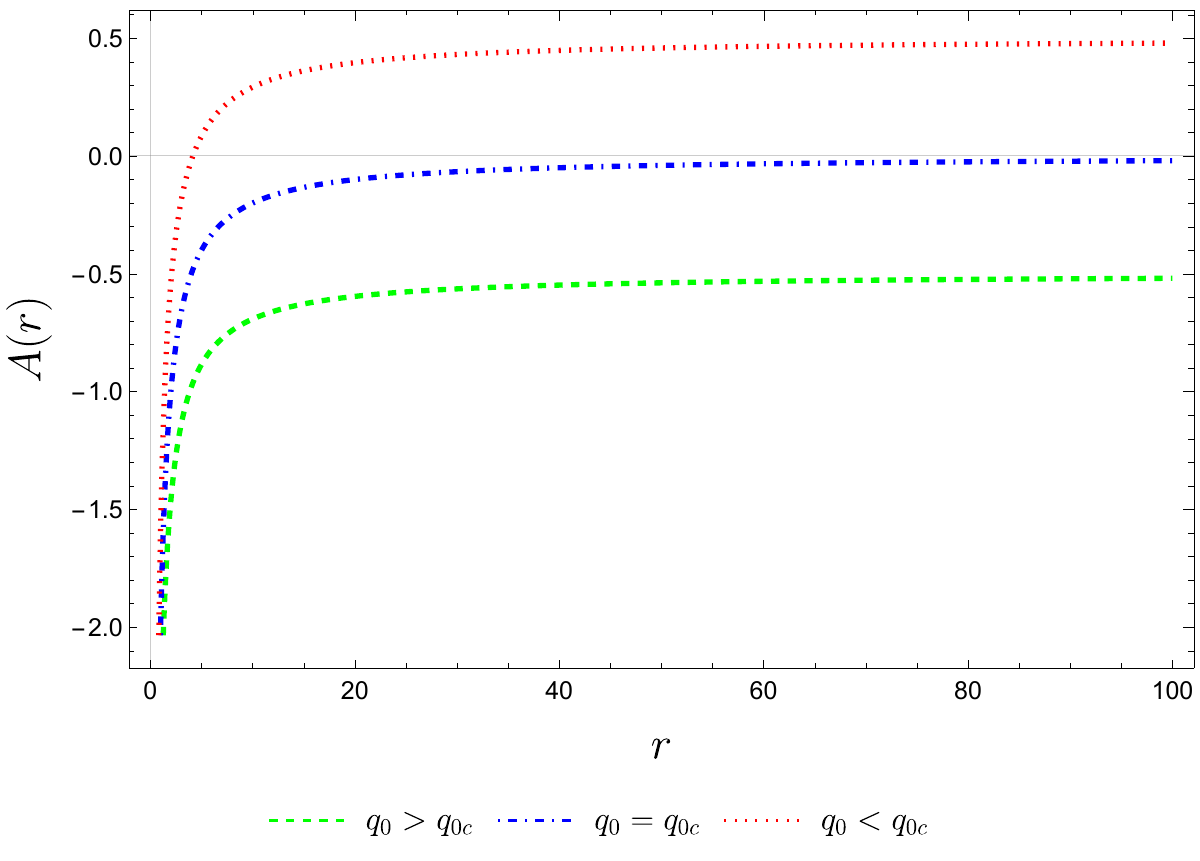}
\caption{The plot depicts the metric function \eqref{A2_BN}, i.e., $A(r)$, for the specific values of $b_0=0.1$ and $M =1.0$.  For $q_0 > 0$ there is no horizon, for $q_{0} = 0$, the curve is negative and tends to zero and for $q_{0} < 0$, there is only one horizon.} 
\label{Axr_BN2}
\end{figure}

We may consider once again Eqs.~\eqref{rH} and~\eqref{der_a} to determine the radius of the event horizon and the critical parameter by numerical solutions. However, for the specific model described by the metric function \eqref{A2_BN}, we did not find an extreme value for $q_{0}$. In Fig. \ref{Axr_BN2}, we depict the behavior of $A(r)$. When $q_0=0$, we observe a curve that is consistently negative and tends towards zero. Conversely, for $q_0> 0$ there is no horizon, while for $q_0 < 0$ there is an horizon. Equation \eqref{A2_BN} shows that the function $A(r)$ tends to $-q_0/2$ in the limit when $r$ tends to infinity. This behavior can be illustrated in Fig.~\eqref{Axr_BN2}, where for $q_0 > 0$ the metric function $A(r)$ stabilizes at a fixed negative value (as shown by the green curve), while for $q_0 < 0$ $A(r)$ tends to a fixed positive value (as shown by the red curve).

In this model the boundary term is given by
\begin{equation}
    \mathbb{B}_Q(r)=\frac{2 (q_0+1) r^4+(3 q_0+4) r^2 r_0^2+2 r_0^4}{r^2 \left(r^2+r_0^2\right)^2}.
    \label{B2_BN}
\end{equation}
Equations \eqref{Q2_BN} and \eqref{B2_BN} satisfy relation \eqref{scalar_Ric2}. 
Although $\mathbb{Q}(r)$ is regular for for all values of $r$ and $q_0$, the combination $\mathbb{Q}(r)-\mathbb{B}_Q(r)=R$ in this model exhibits a singularity in the limit of $r\to0$.

We also find that the quantities ${\cal L}_{\rm NLED}$ and ${\cal L}_F$ take the following forms
\begin{eqnarray}
{\cal L}_{\rm NLED}(r)=-\frac{f(r)}{2\kappa^{2}} 
	-\frac{1}{8\kappa^{2}}\left[\frac{q_{0}r^{3}r_{0}^{2}\left(r^{2}+r_{0}^{2}\right)f'(r)}{(q_{0}+1)r^{6}+(2q_{0}+3)r^{4}r_{0}^{2}+3r^{2}r_{0}^{4}+r_{0}^{6}}\right],
\label{L_BN2}
\end{eqnarray}
and
\begin{equation}
{\cal L}_F(r)=\frac{r^{5}\left(r^{2}+r_{0}^{2}\right)f'(r)\Bigl[(q_{0}+2)r^{4}+4r^{2}r_{0}^{2}+2r_{0}^{4}\Bigr]}{8\kappa^{2}q^{2}\left[(q_{0}+1)r^{6}+(2q_{0}+3)r^{4}r_{0}^{2}+3r^{2}r_{0}^{4}+r_{0}^{6}\right]},
\label{LF_BN2}
\end{equation}
respectively.

Solving the consistency condition~\eqref{RC}, we deduce
\begin{eqnarray}
f(r)=\frac{1}{4}f_{0}\left\{ -\frac{eq_{0}}{r_{0}^{2}}\,
\text{Ei}\left(-\frac{r^{2}}{r_{0}^{2}}-1\right)
-\frac{2}{r_{0}^{2}} \,\text{Ei}\left(-\frac{r^{2}}{r_{0}^{2}}\right)
-\frac{e^{-\frac{r^{2}}{r_{0}^{2}}}\Bigl[(q_{0}+2)r^{4}+2(q_{0}+2)r^{2}r_{0}^{2}+2r_{0}^{4}\Bigr]}{r^{2}\left(r^{2}+r_{0}^{2}\right)^{2}}\right\} +f_{1},
\end{eqnarray}
where $f_0$ and $f_1$ are the integration constants, and ${\rm Ei}(\omega$) is the exponential integral function
\begin{equation}
    \text{Ei}\left(w\right)=\int_{-w}^{\infty}\frac{e^{-t}}{t}dt.
\end{equation}

With the function $f(r)$ at our disposal, the quantities are now expressed in the following form:
\begin{equation}
    f_Q(r)=\frac{1}{4} e^{-\frac{r^2}{r_0^2}}\,,
\end{equation}
\begin{eqnarray} 
 {\cal L}_{\rm NLED}(r)= 
    \frac{1}{8\kappa^{2}}\left\{ \frac{eq_{0}}{r_{0}^{2}}\,
\text{Ei}\left(-\frac{r^{2}}{r_{0}^{2}}-1\right)
+\frac{2}{r_{0}^{2}}\,\text{Ei}\left(-\frac{r^{2}}{r_{0}^{2}}\right)
 +\frac{e^{-\frac{r^{2}}{r_{0}^{2}}}\Bigl[(q_{0}+2)r^{2}+2r_{0}^{2}\Bigr]}{r^{2}\left(r^{2}+r_{0}^{2}\right)}\right\}\,,\label{L2_BN2}
\end{eqnarray}
\begin{eqnarray}
{\cal L}_F(r)=\frac{e^{-\frac{r^{2}}{r_{0}^{2}}}\left[(q_{0}+2)r^{6}+4r^{4}r_{0}^{2}+2r^{2}r_{0}^{4}\right]}{8\kappa^{2}q^{2}\left(r^{2}+r_{0}^{2}\right)^{2}},
\label{LF2_BN2}
\end{eqnarray}
where Eqs.~\eqref{L2_BN2} and~\eqref{LF2_BN2} satisfy the condition~\eqref{RC}.

We can also represent the Lagrangian density as follows
\begin{eqnarray}
 {\cal L}  _{\rm NLED}(F) =  \frac{1}{8\kappa^{2}}\left[
 \frac{e^{-\frac{q}{\sqrt{2F}r_{0}^{2}}}\left(2\sqrt{F}q(q_{0}+2)+4\sqrt{2}Fr_{0}^{2}\right)}{q\left(2\sqrt{F}r_{0}^{2}+\sqrt{2}q\right)}
+\frac{eq_{0}}{r_{0}^{2}}\text{Ei}\left(-\frac{q}{\sqrt{2F}r_{0}^{2}}-1\right)+\frac{2}{r_{0}^{2}}\text{Ei}\left(-\frac{q}{\sqrt{2F}r_{0}^{2}}\right)
 \right].
 \label{LxF3_BN2}
\end{eqnarray}

\begin{figure}[h]
\centering
\includegraphics[scale=0.4]{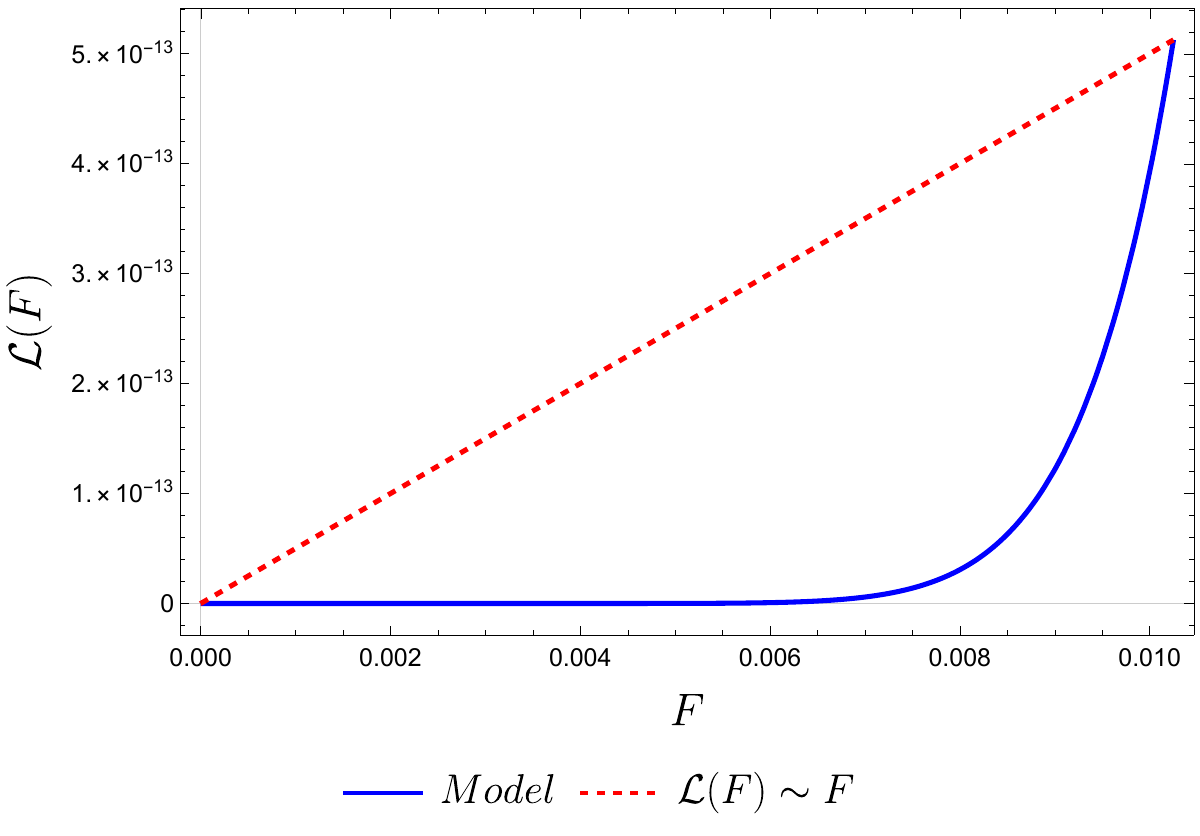}
\caption{Graphic representation of ${\cal L}_{\rm NLED}(F)$, described by expression~\eqref{LxF3_BN2}, with respect to $F$. Where we consider $q=0.7$ and $r_0=0.5$.  The blue curve represents the behavior of our model for $F\ll1$, while the dashed red line represents the linear case.} 
\label{iLxF2_BN2}
\end{figure}

The asymptotic limits of the Lagrangian \eqref{LxF3_BN2} for $F\gg1$ are
\begin{align}
    {\cal L} _{\rm NLED} (F)\sim &-0.219\frac{eq_{0}}{8\kappa^{2}r_{0}^{2}}+\frac{\sqrt{F}}{2\sqrt{2}\kappa^{2}q}+\frac{-\ln(F)+q_{0}+2\gamma-2}{8\kappa^{2}r_{0}^{2}}-\frac{\sqrt{\frac{1}{F}}q\left(2q_{0}+1\right)}{8\left(\sqrt{2}\kappa^{2}r_{0}^{4}\right)}+\frac{q^{2}(15q_{0}+1)}{96F\kappa^{2}r_{0}^{6}},
\end{align}
where $\gamma$ is the Euler constant. The numerical value 0.219 came from an approximation of the integral exponential function.
We have constructed a plot to demonstrate the asymptotic limit of Lagrangian \eqref{LxF3_BN2} for $F\ll1$. To illustrate this behavior, we compared the curve of our model with a Lagrangian proportional to $F$, i.e. the linear case, represented by the blue curve and the dashed red line, respectively, as depicted in Fig. \ref{iLxF2_BN2}.

We see that we cannot recover the linear case if we take the asymptotic limits.

\begin{figure}[h]
\centering
\includegraphics[scale=0.4]{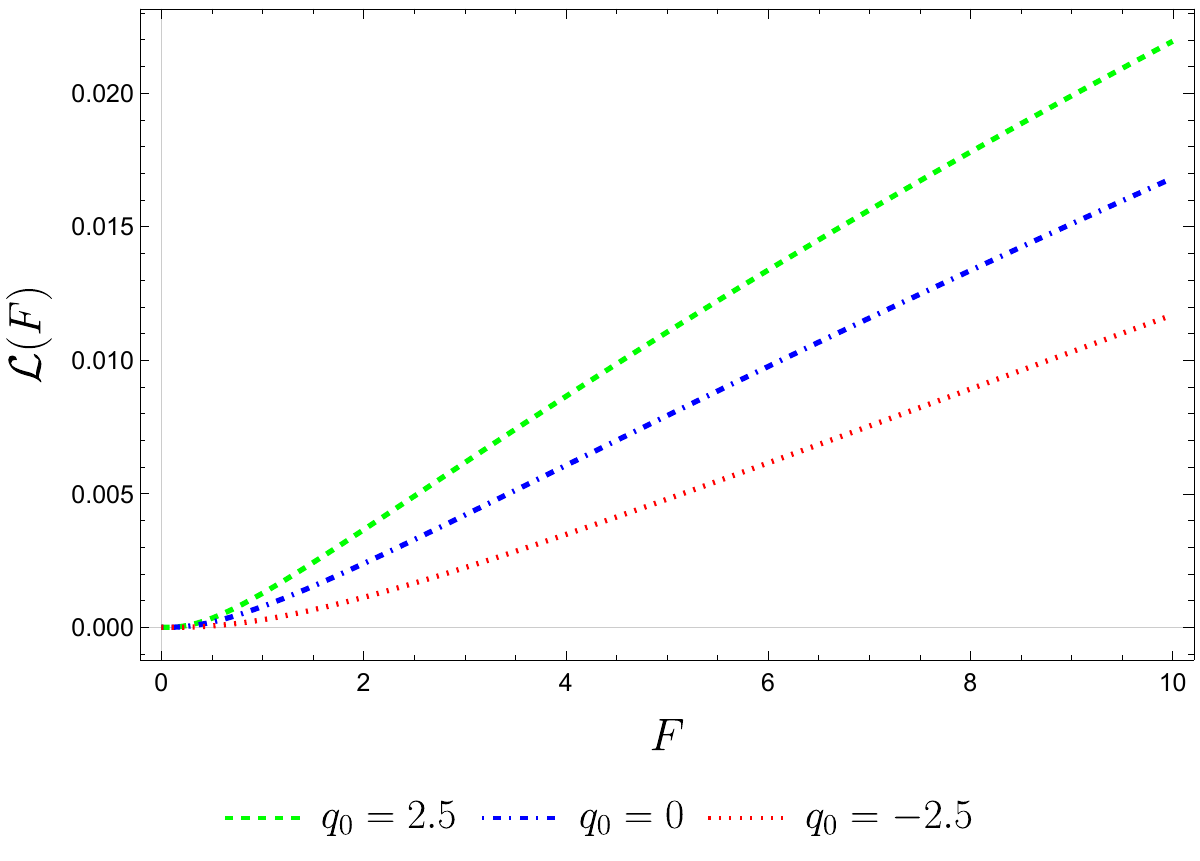}
\caption{Graphic representation of ${\cal L}_{\rm NLED}(F)$, described by expression~\eqref{LxF3_BN2}, with respect to $F$. Where we consider $q=0.7$ and $r_0=0.5$.} 
\label{LxF_BN2}
\end{figure}

The behavior of Lagrangian~\eqref{LxF3_BN2} for three different scenarios of $q_0$ ($q_0 > 0$, $q_0 = 0$, $q_0 < 0$) is described in Fig.~\ref{LxF_BN2}.

In Fig.~\ref{LxF_BN2}, we see that we get different curves of ${\cal L}_{\rm NLED}(F)$ (Eq. \eqref{LxF3_BN2}) for different values of $q_0$ if we keep the values of $q$ and $r_0$ constant.
Similarly, we could hold the values of $q_0$ and $q$ constant and vary only the value of $r_0$ or alternatively hold $r_0$ and $q_0$ constant and vary $q$. From the graph we can also see that the Lagrangian ${\cal L}_{\rm NLED}(F)$ (Eq. \eqref{LxF3_BN2}) is a monotonically increasing function. This could be a characteristic of a solution in nonlinear magnetically charged electrodynamics.

\subsection{Third model of black hole}\label{BN3}

In constructing the third model, we assume again the symmetry provided by~\eqref{Symmetry2}, and consider the following regular non-metricity scalar
\begin{equation}
    \mathbb{Q}(r)=\frac{q_0}{(r+r_0)^2}.\label{Q3_BN}
\end{equation}
Thus, from Eq.~\eqref{Q_BN}, we have
\begin{equation}
   -\frac{2 \Bigl(r A'(r)+A(r)\Bigr)}{r^2}=\frac{q_0}{(r+r_0)^2},
\end{equation}
which yields the following solution 
\begin{equation}
    A(r)=-\frac{2M}{r}-\frac{q_{0}}{2r}\left[r-\frac{r_{0}^{2}}{r+r_{0}}-2r_{0}\ln(r+r_{0})\right].\label{A3_BN}
\end{equation}
The metric function~\eqref{A3_BN} diverges in the limit of $r\to0$, but becomes asymptotically flat at spatial infinity, $r\to\infty$.

\begin{figure}[h]
\centering
\includegraphics[scale=0.4]{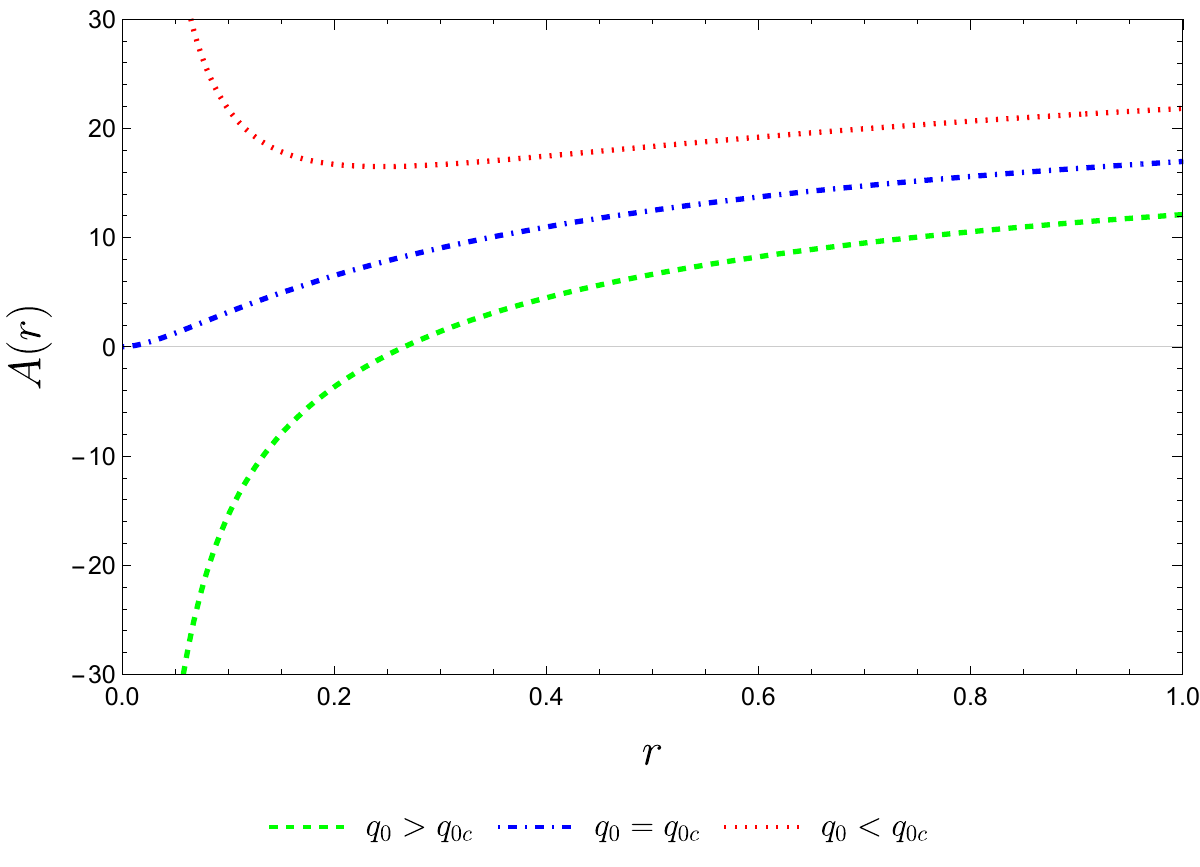}
\caption{The plot depicts the metric function \eqref{A1_BN}, i.e., $A(r)$, for the specific values of $r_0=0.1$ and $M =5.0$, the critical value of the constant $q_0$, given by $q_{0c} = -55.475$.  For $q_0 > q_{0c}$ there is only one horizon, for $q_0 = q_{0c}$, there is  light-type  singularity and for $q_0 < q_{0c}$, there is no horizon.} 
\label{Axr_BN3}
\end{figure}

Here, too, we analyze the presence of horizons and the critical parameter $q_{0c}$ using Eqs.~\eqref{rH} and \eqref{der_a}, based on numerical solutions. Using these equations and the values for the constants $r_0=0.1$ and $M=1$, we find the critical parameter value $q_{0c}=-55.475$. Under these assumptions, we plot the metric function \eqref{A3_BN} in Fig. \ref{Axr_BN3} for three scenarios: $q_0 > q_{0c}$, $q_0 = q_{0c}$, and $q_0 < q_{0c}$. When $q_0$ exceeds the critical value, $q_0 > q_{0c}$, we observe the presence of an horizon.  For $q_0 = q_{0c}$, we observe a  light-type  singularity, and if $q_0 < q_{0c}$, no horizon is formed.

In this scenario, the boundary term is given by
\begin{align}
  \mathbb{B}_Q(r)=&\frac{2 (q_0+1) r^3+3 (q_0+2) r^2 r_0+6 r r_0^2+2 r_0^3}{r^2 (r+r_0)^3},\label{B3_BN}
\end{align}
so that the scalar~\eqref{Q3_BN} and  boundary term~\eqref{B3_BN} satisfy the condition~\eqref{scalar_Ric2}. In this model, we also find that the combination   $\mathbb{Q}(r)-\mathbb{B}_Q(r)$  is singular in the limit of $r\to0$.

Now, Eqs. \eqref{L2_BN} and \eqref{LF2_BN} are expressed as
\begin{align}
 {\cal L}_{\rm NLED}(r)&=-\frac{1}{2\kappa^{2}}\left[\frac{q_{0}r^{3}r_{0}(r+r_{0})f'(r)}{4(q_{0}+1)r^{4}+(7q_{0}+16)r^{3}r_{0}+24r^{2}r_{0}^{2}+16rr_{0}^{3}+4r_{0}^{4}}+f(r)\right],\label{L_BN3} \\
 {\cal L}_F(r)&=\frac{r^{5}(r+r_{0})f'(r)\Bigl[(q_{0}+2)r^{3}+6r^{2}r_{0}+6rr_{0}^{2}+2r_{0}^{3}\Bigr]}{2\kappa^{2}q^{2}\Bigl[4(q_{0}+1)r^{4}+(7q_{0}+16)r^{3}r_{0}+24r^{2}r_{0}^{2}+16rr_{0}^{3}+4r_{0}^{4}\Bigr]}.\label{LF_BN3}
\end{align}

Using Eqs. \eqref{L_BN3} and \eqref{LF_BN3} in the relation \eqref{RC}, allows us to determine 
\begin{equation}
    f(r)=\frac{f_{0}}{r_{0}^{2}}\left[4e^{2}q_{0}\text{Ei}\left(-\frac{2(r+r_{0})}{r_{0}}\right)+8\text{Ei}\left(-\frac{2r}{r_{0}}\right)+\frac{r_{0}e^{-\frac{2r}{r_{0}}}\left(2(q_{0}+2)r^{4}+(3q_{0}+10)r^{3}r_{0}+6r^{2}r_{0}^{2}-2rr_{0}^{3}-2r_{0}^{4}\right)}{r^{2}(r+r_{0})^{3}}\right]+f_1\,,
\end{equation}
where $f_0$ and $f_1$ are integration constants, so that Eqs. \eqref{L_BN3} and \eqref{LF_BN3}  are given by
\begin{equation}
 {\cal L}_{\rm NLED} (r)=\frac{1}{2\kappa^{2}r_{0}^{2}}\left\{ -4e^{2}q_{0}\text{Ei}\left(-\frac{2(r+r_{0})}{r_{0}}\right)-8\text{Ei}\left(-\frac{2r}{r_{0}}\right)+\frac{r_{0}e^{-\frac{2r}{r_{0}}}\left[-2(q_{0}+2)r^{3}-(q_{0}+6)r^{2}r_{0}+2r_{0}^{3}\right]}{r^{2}(r+r_{0})^{2}}\right\} ,\label{L2_BN3} 
 \end{equation}
 \begin{equation}
 {\cal L}_F (r) =\frac{r^{2}e^{-\frac{2r}{r_{0}}}\Bigl[(q_{0}+2)r^{3}+6r^{2}r_{0}+6rr_{0}^{2}+2r_{0}^{3}\Bigr]}{2\kappa^{2}q^{2}(r+r_{0})^{3}},
 \label{LF2_BN3}
\end{equation}
respectively.

\begin{figure}[h]
\centering
\includegraphics[scale=0.4]{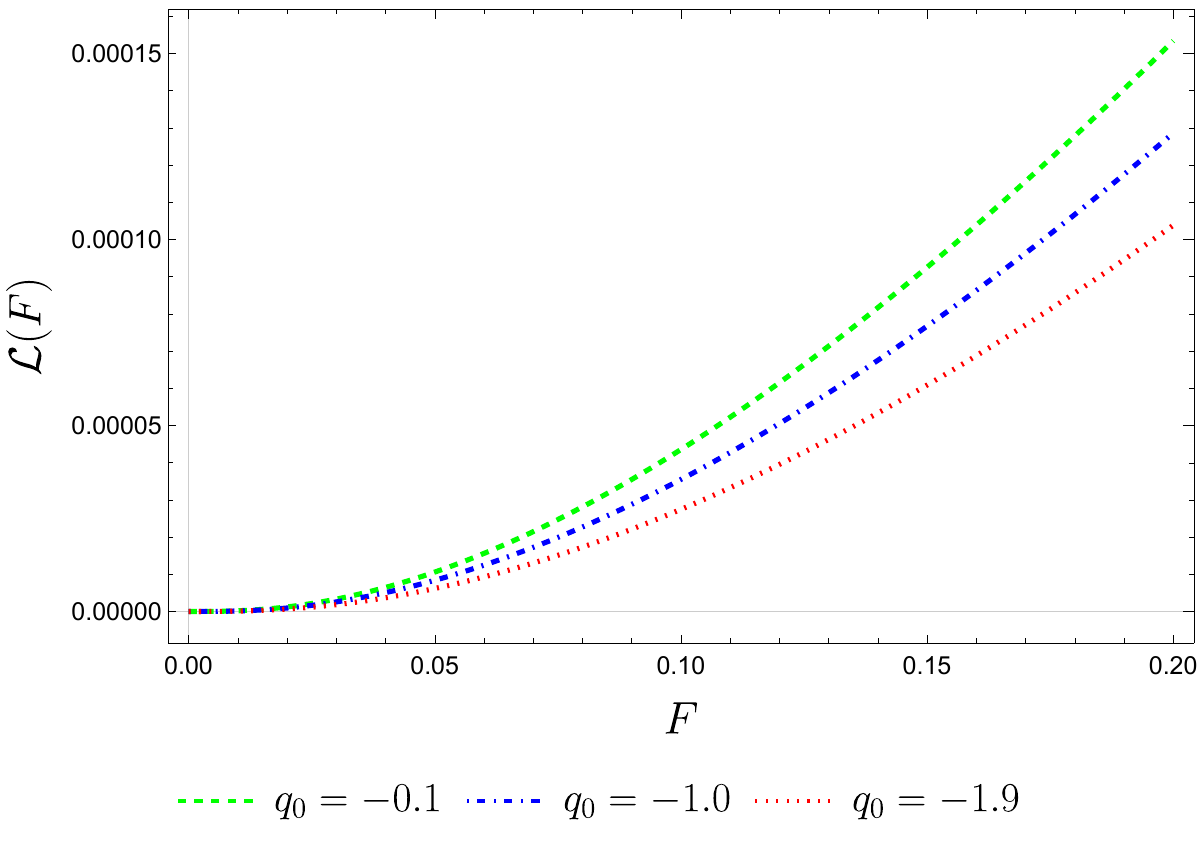}
\caption{The plot depicts ${\cal L}_{\rm NLED}(F)$, described by Eq.~\eqref{BN3_LxF}, for the values $q_0=-0.1$, $q_0=-1.0$ and $q_0=-1.9$.} 
\label{LxF_BN3}
\end{figure}

Finally, using Eq.~\eqref{F}, we derive the expression for $r(F)$, so that
\begin{align}
     {\cal L}_{\rm NLED} (F)= & \,\,\,\frac{e^{-\frac{2^{3/4} \sqrt{q}}{\sqrt[4]{F} r_0}} \left(-2\ 2^{3/4} \sqrt[4]{F} q^{3/2} (q_0+2)-2 \sqrt{F} q (q_0+6) r_0+4 \sqrt{2} F r_0^3\right)}{\kappa ^2 q r_0 \left(2 \sqrt[4]{F} r_0+2^{3/4} \sqrt{q}\right)^2}
     \nonumber\\
   & \,\, -\frac{2 }{\kappa ^2 r_0^2} \left[e^2 q_0 \text{Ei}\left(-\frac{2^{3/4} \sqrt{q}}{\sqrt[4]{F} r_0}-2\right)+2 \text{Ei}\left(-\frac{2^{3/4} \sqrt{q}}{\sqrt[4]{F} r_0}\right)\right],
   \label{BN3_LxF}
\end{align}
which is depicted in in Fig.~\ref{LxF_BN3} for different values of $q_0$. We note that ${\cal L}_{\rm NLED} (F)$ is non-linear and has the properties of an increasing bijective function. If we fix the values of $r_0$ and $q$, we find different models for varying values of $q_0$.

\begin{figure}[h]
\centering
\includegraphics[scale=0.4]{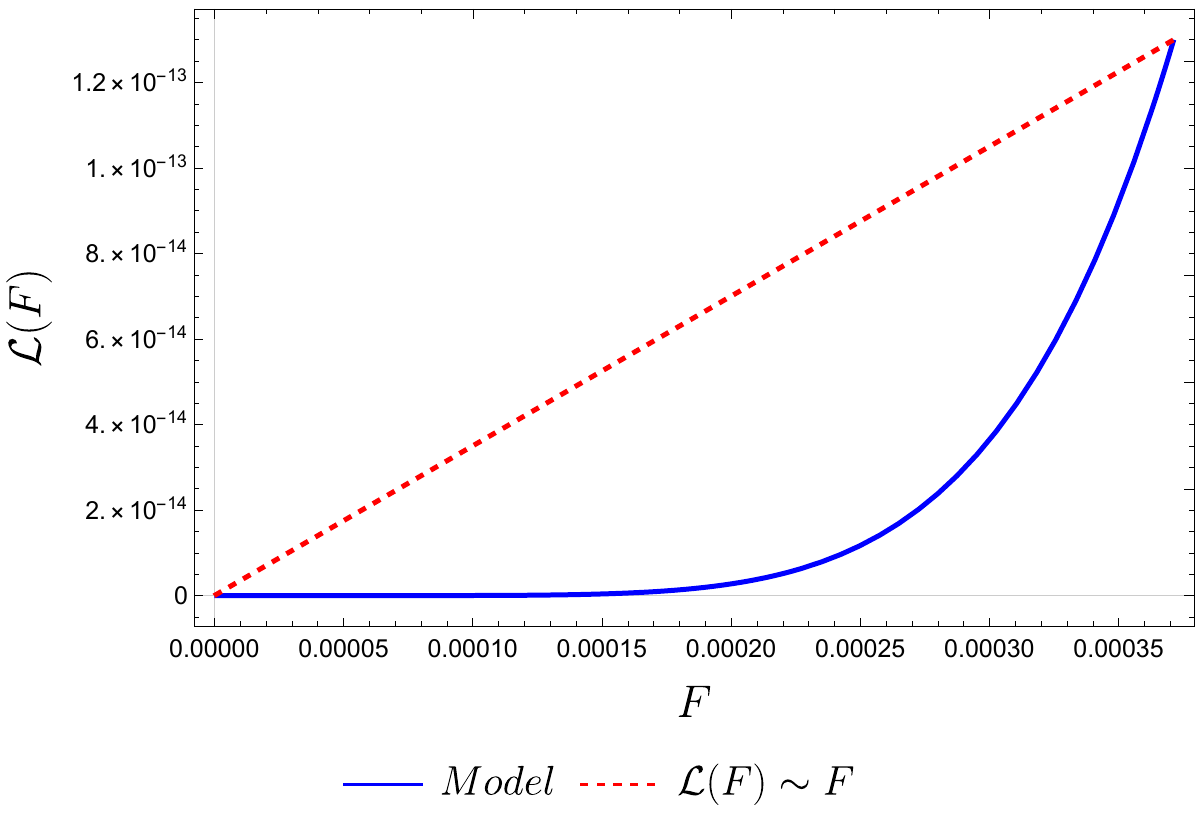}
\caption{Graphic representation of ${\cal L}_{\rm NLED}(F)$, described by expression~\eqref{BN3_LxF}, with respect to $F$. Where we consider $q=0.7$, $q_0=-1.0$ and $r_0=0.5$. The blue curve represents the behavior of our model for $F\ll1$, while the dashed red line represents the linear case.} 
\label{iLxF2_BN3}
\end{figure}

The asymptotic limits of the Lagrangian \eqref{LxF_BN3} for $F\gg1$ are
\begin{align}
    {\cal L} _{\rm NLED} (F)\sim &0.01\frac{e^{2}q_{0}}{\kappa^{2}r_{0}^{2}}+\frac{\sqrt{2}\sqrt{F}}{\kappa^{2}q}-\frac{4\sqrt[4]{2}\sqrt[4]{F}}{\kappa^{2}\sqrt{q}r_{0}}+\frac{\sqrt[4]{\frac{1}{F}}\sqrt{q}(3q_{0}+8)}{3\sqrt[4]{2}\kappa^{2}r_{0}^{3}}+\frac{2\ln(F)-q_{0}-8\gamma+12}{2\kappa^{2}r_{0}^{2}}\nonumber\\&-\frac{2\left(\frac{1}{F}\right)^{3/4}\left(15\sqrt[4]{2}q^{3/2}q_{0}-2\sqrt[4]{2}q^{3/2}\right)}{45\left(\kappa^{2}r_{0}^{5}\right)}+\frac{375q^{2}q_{0}-4q^{2}}{180F\kappa^{2}r_{0}^{6}}-\frac{\sqrt{\frac{1}{F}}(q(3q_{0}+4))}{6\left(\sqrt{2}\kappa^{2}r_{0}^{4}\right)},
\end{align}
where $\gamma$ is the Euler constant. The numerical value 0.01 came from an approximation of the integral exponential function.
We illustrate again the behavior described by the Lagrangian of our model, Eq. \eqref{BN3_LxF}, with a Lagrangian that is proportional to the linear case, represented by the blue curve and the dashed red line, respectively, as described in Fig. \ref{iLxF2_BN3}.

\section{Regular Black Hole solutions in $f(\mathbb{Q,B_Q})$ gravity} \label{sec5}

In this section, we explore solutions for regular black holes, taking into account the metric functions of the Bardeen \cite{Bardeen} and Culetu solutions \cite{Culetu:2013fsa,Culetu:2014lca}. To obtain these solutions, we use again the symmetry imposed in Eq. \eqref{Symmetry2}, and the line element~\eqref{m_bn}.

\subsection{Bardeen-type solutions }\label{subsec5A}

For the first solution, consider the metric function described by Bardeen's model \cite{Bardeen}, given by
\begin{equation}
    A(r)=1-\frac{2Mr^{2}}{(q^{2}+r^{2})^{3/2}},
    \label{A_Bardeen}
\end{equation}
so that Eqs.~\eqref{B} and~\eqref{Q} take the following forms:
\begin{eqnarray}
&& \mathbb{B}_Q(r)= \frac{6 M q^2 r^2}{\left(q^2+r^2\right)^{7/2}}+\frac{36 M q^4}{\left(q^2+r^2\right)^{7/2}}-\frac{2}{r^2},\label{B_Bardeen} \\
&&    \mathbb{Q}(r)=\frac{12 M q^2}{\left(q^2+r^2\right)^{5/2}}-\frac{2}{r^2},
	\label{Q_Bardeen}
\end{eqnarray}
respectively.
We verify that in the limit of $r\to0$ the combination of $\mathbb{Q}(r)-\mathbb{B}_Q(r)$ from Eqs. \eqref{B_Bardeen} and \eqref{Q_Bardeen} is regular, so that the spacetime geometry is regular.

The following electromagnetic relations are given by
\begin{equation}
 {\cal L}_{\rm NLED} (r) =\frac{1}{2 \kappa ^2} \Biggl[\frac{3 M q^2 r^3 \left(3 r^2-2 q^2\right) f'(r)}{\left(q^2+r^2\right)^{7/2}-15 M q^2 r^4}-2 f(r)\Biggr], \label{L_Bardeen_RBN}
\end{equation}
\begin{equation}
 {\cal L}_F (r) =\frac{15 M r^9 f'(r)}{4 \kappa ^2 \left[\left(q^2+r^2\right)^{7/2}-15 M q^2 r^4\right]},\label{LF_Bardeen_RBN}
\end{equation}
so that from the consistency relation~\eqref{RC}, we obtain the following expression:
\begin{eqnarray}
f(r)=f_{0}\int\left(\int\exp\left(\zeta\right)dr\right)dr+f_{1},\label{f_Bardeen}
\end{eqnarray}
where $f_0$ and $f_1$ are constants, to simplify this expression and subsequent ones, we define
\begin{eqnarray}
&& \zeta= -\frac{1}{3Mq^{2}r^{3}\left(2q^{2}-3r^{2}\right)\left(\left(q^{2}+r^{2}\right)^{7/2}-15Mq^{2}r^{4}\right)}\Biggl[2q^{14}+14q^{12}r^{2}+2r^{14}+
	\nonumber\\
&&+q^{8}\left(70r^{6}-123Mr^{4}\sqrt{q^{2}+r^{2}}\right)+3q^{4}r^{8}\left(345M^{2}-101M\sqrt{q^{2}+r^{2}}+14r^{2}\right)+2q^{2}r^{10}\left(7r^{2}-36M\sqrt{q^{2}+r^{2}}\right)
	\nonumber\\ 
&& \phantom{f(r)=}+6q^{10}\left(3Mr^{2}\sqrt{q^{2}+r^{2}}+7r^{4}\right)+2q^{6}r^{6}\left(45M^{2}-186M\sqrt{q^{2}+r^{2}}+35r^{2}\right)\Biggl].
\end{eqnarray}
From Eq.~\eqref{f_Bardeen}, we obtain the derivative of $f_Q(r)$, given by
\begin{eqnarray}
 &&  f_{Q}(r)=\frac{r^3 \left(q^2+r^2\right)^{7/2} }{4 \left(\left(q^2+r^2\right)^{7/2}-15 M q^2 r^4\right)}\int \exp\left(\zeta\right) \, dr.\nonumber \\
\label{fB_Bardeen}
\end{eqnarray}

\begin{figure*}[h]
\centering
\includegraphics[scale=0.5]{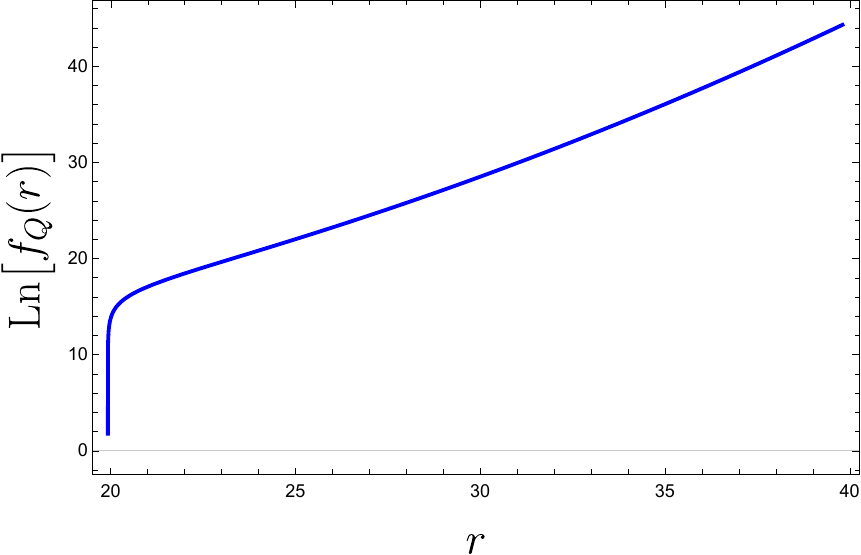}
\quad\quad\quad\quad\quad\quad
\includegraphics[scale=0.35]{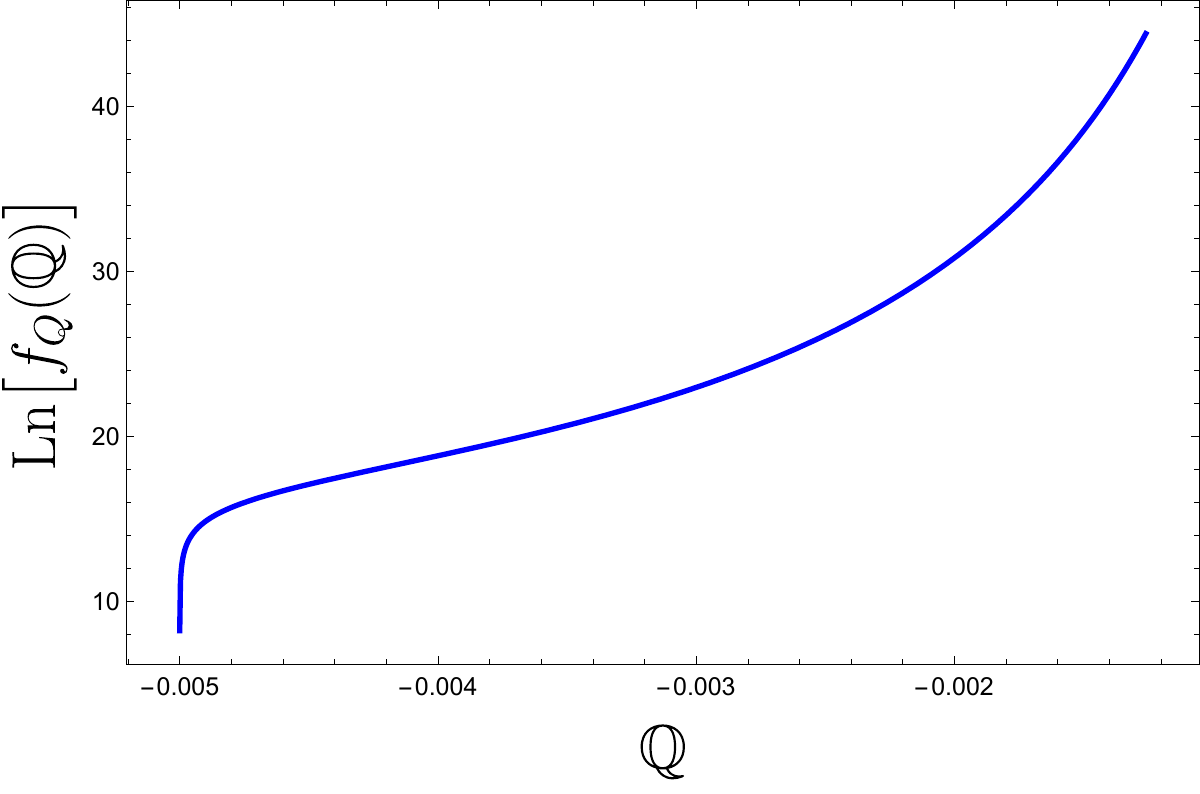}
\caption{The left plot depicts the graphical representation of the expression $\textrm{Ln}[f_Q(r)]$, described by Eq.~\eqref{fB_Bardeen}, with respect to the coordinate $r$. The right plot represents the graphic representation of $\textrm{Ln}[f_Q(\mathbb{Q})]$ with respect to the boundary term $\mathbb{Q}$. We have considered $M=10$ and $q=1$.} 
\label{Fig1_Bardeen}
\end{figure*}

We depict the linearity of the function $f_Q(r)$, described by  Eq.~\eqref{fB_Bardeen} , with respect to the non-metricity scalar in two ways in Fig. \ref{Fig1_Bardeen}. On the left, the graph illustrates the variation of the natural logarithm of $f_Q(r)$ with respect to $r$, highlighting a non-linear dependence on $\mathbb{Q}$. On the right, it is evident that the function $f_Q(\mathbb{Q})$ is non-linear with respect to $\mathbb{Q}$. Furthermore, we observe that the function $f(\mathbb{Q},\mathbb{B}_Q)$ is also non-linear with respect to $\mathbb{Q}$.

With these expressions we find that the Lagrangian density is now given by
\begin{eqnarray}
  {\cal L} _{\rm NLED}(r)=\frac{1}{4\kappa^{2}}\left[\frac{M\left(9q^{2}r^{5}-6q^{4}r^{3}\right)}{\left(q^{2}+r^{2}\right)^{7/2}-15Mq^{2}r^{4}}\int \exp\left(\zeta\right)\,dr-2f(r)\right],\label{L2_Bardeen}
\end{eqnarray}
where $f(r)$ in the  above equation is given by Eq.~\eqref{f_Bardeen}, and ${\cal L}_F(r)$ is
\begin{equation}
  {\cal L}_F(r) = \frac{15Mr^{9}}{4\kappa^{2}\left[\left(q^{2}+r^{2}\right)^{7/2}-15Mq^{2}r^{4}\right]}\int\exp\left(\zeta\right)dr.\label{LF2_Bardeen}
\end{equation}

Equation \eqref{LF2_Bardeen} is depicted in Fig. \ref{Fig2_Bardeen}, where we verify that ${\cal L}_F(F)$ is a non-linear and bijective function with respect to $F$.

\begin{figure}[h]
\centering
\includegraphics[scale=0.55]{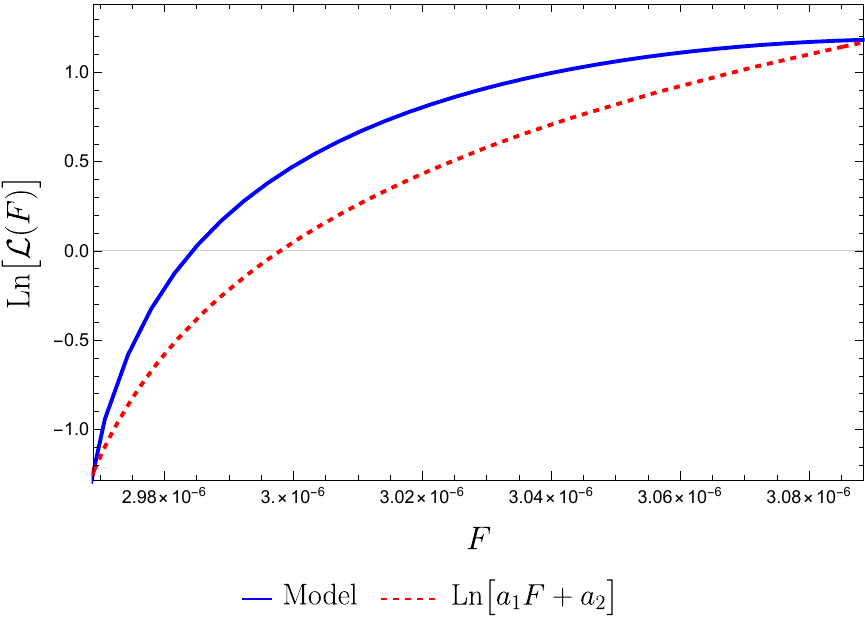}
\caption{Graphic representation of ${\cal L}_{\rm NLED}(F)$, described by expression~\eqref{L2_Bardeen}, with respect to $F$. We consider here  $M=10$, $q=1$, $a_1=2.45\times 10^7$ and $a_2=-72.45$ .  The blue curve represents the behavior of our model for $F\ll1$, while the dashed red line represents the linear case, both described by the logarithmic function.} 
\label{LxF_Bard2}
\end{figure}

\begin{figure}[h]
\centering
\includegraphics[scale=0.4]{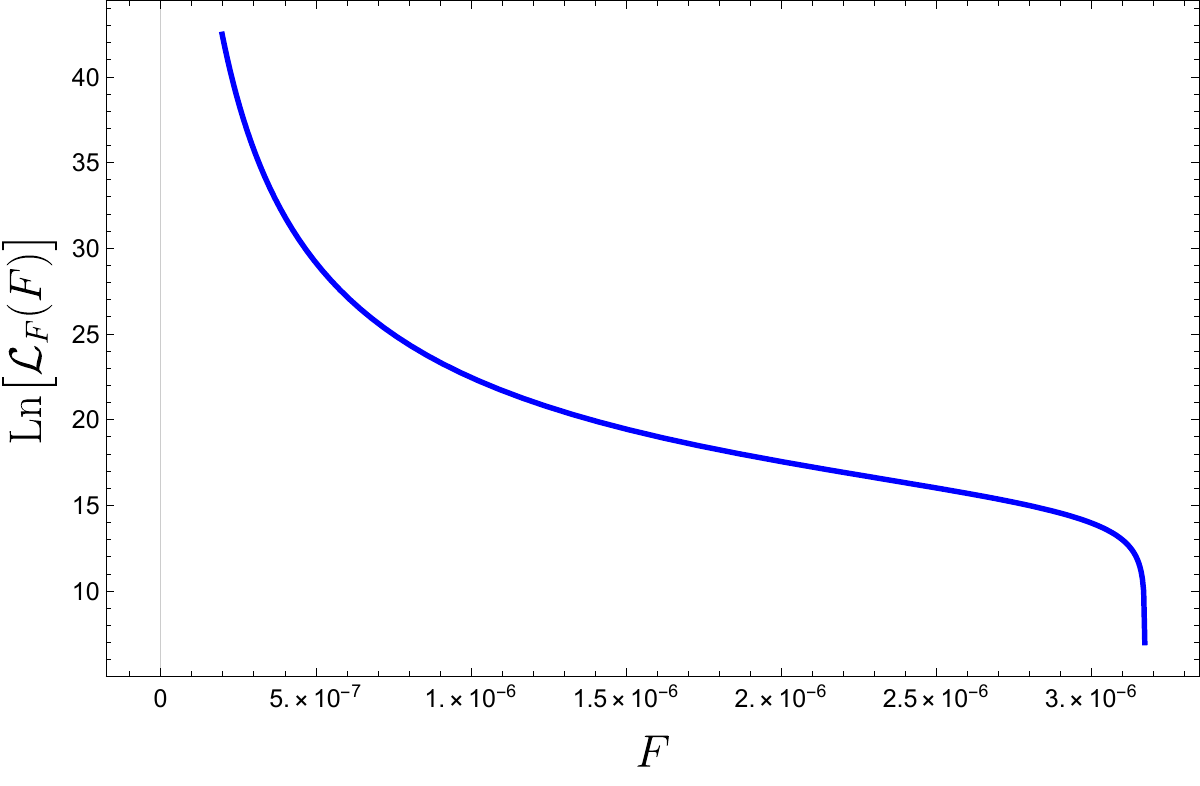}
\caption{Graphic representation of ${\cal L}_F(F)$, described by Eq.~\eqref{LF2_Bardeen}, with respect to $F$, where we have considered $M=10$ and $q=1$.} 
\label{Fig2_Bardeen}
\end{figure}

In Figure \ref{LxF_Bard2} we show a comparison of the behavior of the Lagrangian obtained in our model~\eqref{L2_Bardeen} with a Lagrangian from the linear case.

\subsection{Culetu-type solutions }\label{subsec5B}

In this second approach, we use the model proposed by Culetu \cite{Culetu:2013fsa,Culetu:2014lca}:
\begin{equation}
    A(r)=1-\frac{2 M }{r} \exp \left(-\frac{q^2}{2 M r}\right),
    \label{A_Culetu}
\end{equation}
so that Eqs. \eqref{B} and \eqref{Q} take the following forms:
\begin{equation}
 \mathbb{B}_Q(r)= \frac{e^{-\frac{q^2}{2 M r}} \left(-4 M r^3 e^{\frac{q^2}{2 M r}}+4 M q^2 r+q^4\right)}{2 M r^5},
 \label{B_Culetu} 
 \end{equation}
\begin{equation}  
 \mathbb{Q}(r)=\frac{2 q^2 e^{-\frac{q^2}{2 M r}}-2 r^2}{r^4},
 \label{Q_Culetu}
\end{equation}
respectively. We verify that, in the limit $r \rightarrow 0$, the combination of $\mathbb{Q}(r) - \mathbb{B}_Q(r)$ from Eqs. \eqref{B_Culetu} and \eqref{Q_Culetu} is regular, rendering a regular spacetime geometry.

We also deduce the following relations
\begin{eqnarray}
 {\cal L}_{\rm NLED}(r)=-\frac{1}{2 \kappa ^2} \Biggl(f(r)-\frac{2 M q^2 r^2 f'(r) \left(q^2-4 M r\right)}{16 M^2 r^4 e^{\frac{q^2}{2 M r}}-32 M^2 q^2 r^2-6 M q^4 r+q^6}\Biggr),\label{L_Culetu_RBN}
\end{eqnarray}
\begin{eqnarray}
{\cal L}_F (r) =\frac{M r^6 f'(r) \left(q^2-8 M r\right)}{\kappa ^2 \left(16 M^2 r^4 e^{\frac{q^2}{2 M r}}-32 M^2 q^2 r^2-6 M q^4 r+q^6\right)},
\label{LF_Culetu_RBN}
\end{eqnarray}
which from the consistency condition~\eqref{RC}, provides the following 
\begin{eqnarray}
&& f(r)=\int\exp\left\{ -\frac{e^{2}q^{4}\text{Ei}\left(\frac{q^{2}}{2Mr}-2\right)+q^{4}\left[\frac{4M}{r}-5\text{Ei}\left(\frac{q^{2}}{2Mr}\right)\right]+8Mre^{\frac{q^{2}}{2Mr}}\left(Mr+q^{2}\right)}{8M^{2}q^{2}}\right\} \times\nonumber\\
&&\phantom{f(r)=}\times\frac{f_{0}\left(q^{2}-4Mr\right)^{2}\left(-32M^{2}q^{2}r^{2}+16M^{2}r^{4}e^{\frac{q^{2}}{2Mr}}-6Mq^{4}r+q^{6}\right)}{r^{5}}dr   +f_1 ,\label{f_Culetu}
\end{eqnarray}
where $f_0$ and $f_1$ are constants.

Unfortunately, we were unable to obtain an analytic solution for the function $f(r)$ directly from the consistency relation. However, we succeeded in determining the function $f_B(r)$, which is described as
\begin{equation}
    f_B(r)=4 M^2 r^2 \left(q^2-4 M r\right)^2 \exp \left\{-\frac{q^4 \left[e^2 \text{Ei}\left(\frac{q^2}{2 M r}-2\right)-5 \text{Ei}\left(\frac{q^2}{2 M r}\right)\right]+8 M r e^{\frac{q^2}{2 M r}} \left(M r+q^2\right)}{8 M^2 q^2}\right\}.\label{fB_Culetu}
\end{equation}

\begin{figure}[h]
\centering
\includegraphics[scale=0.54]{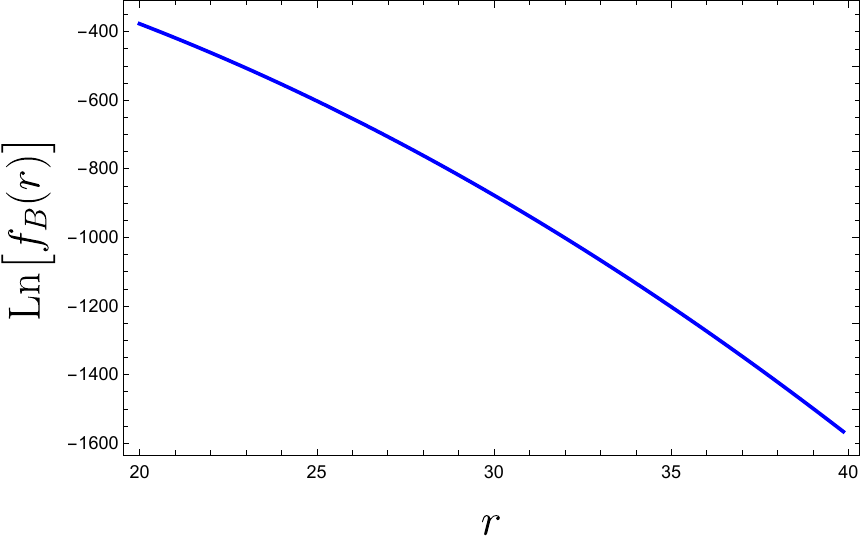}
\quad\quad\quad
\includegraphics[scale=0.385]{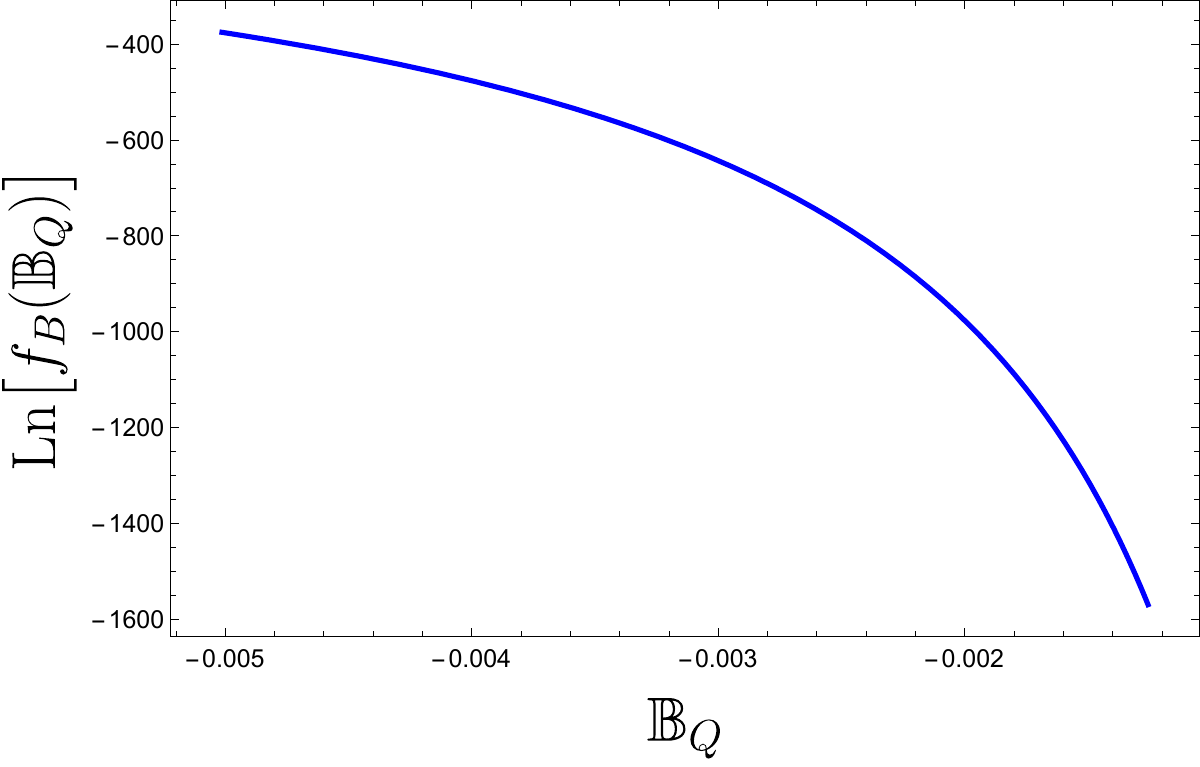}
\caption{The left plot depicts $\textrm{Ln}[f_B(r)]$, described by expression~\eqref{fB_Culetu}, with respect to the radial coordinate $r$.  The right plot represents $\textrm{Ln}[f_B(r)]$ with respect to the boundary term $\mathbb{B}_Q$. 
We consider $M=10$ and $q=1$.} 
\label{Fig1_Culetu}
\end{figure}

The qualitative behavior of Eq.~\eqref{fB_Culetu} with respect to the radial coordinate $r$ is depicted in the left plot of Fig.~\ref{Fig1_Culetu},  where the function  $f_B(r)$ is non-linear in relation to $r$, indicating non-linearity with respect to the boundary term $\mathbb{B}$. This can be directly verified in the right plot of Fig.~\ref{Fig1_Culetu}, where the function  $f_B$  is clearly non-linear. Therefore, we conclude that the function  $f(\mathbb{Q},\mathbb{B})$ must be non-linear in the boundary term $\mathbb{B}_Q$.


Taking into account Eqs.~\eqref{f_Culetu} and~\eqref{fB_Culetu}, we find that the Lagrangian density is now given by
\begin{eqnarray}
    {\cal L} _{\rm NLED}(r)&&=-\frac{1}{2\kappa^{2}}\left[-\exp\left\{ -\frac{q^{2}\left[-5r\text{Ei}\left(\frac{q^{2}}{2Mr}\right)+e^{2}r\text{Ei}\left(\frac{q^{2}}{2Mr}-2\right)+4M\right]}{8M^{2}r}-\frac{re^{\frac{q^{2}}{2Mr}}\left(Mr+q^{2}\right)}{Mq^{2}}\right\} \times\right.\nonumber\\&&\left.\times\frac{2Mq^{2}\left(q^{2}-4Mr\right)^{3}}{r^{3}}+f(r)\right]\label{L_Culetu}
\end{eqnarray}
and its derivative is
\begin{equation}
  {\cal L}_F (r) = \frac{M r \left(q^2-8 M r\right) \left(q^2-4 M r\right)^2 }{\kappa ^2}\exp \left[-\frac{q^2 \left(-5 r \text{Ei}\left(\frac{q^2}{2 M r}\right)+e^2 r \text{Ei}\left(\frac{q^2}{2 M r}-2\right)+4 M\right)}{8 M^2 r}-\frac{r e^{\frac{q^2}{2 M r}} \left(M r+q^2\right)}{M q^2}\right].\label{LF2_Culetu}
\end{equation}

\begin{figure}[!h]
\centering
\includegraphics[scale=0.4]{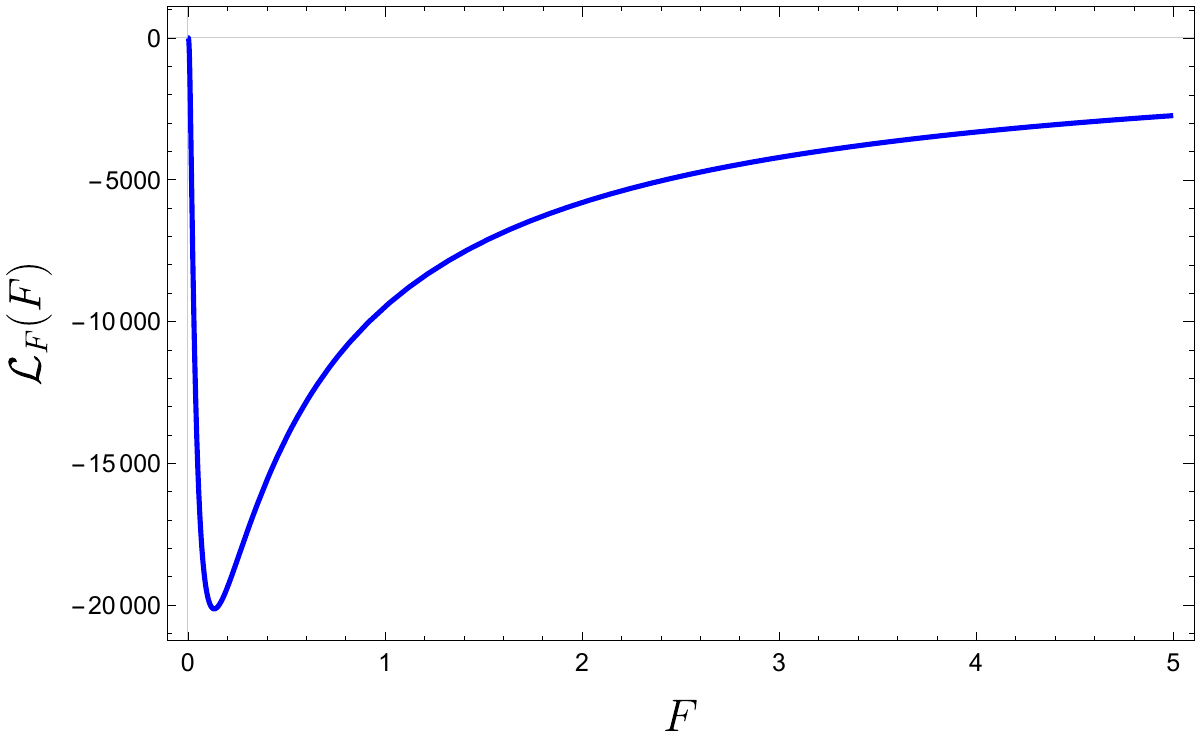}
\caption{Graphical representation of the derivative of the electromagnetic Lagrangian ${\cal L} _F(F)$, described by expression~\eqref{LF3_Culetu}, as a function of $F$. We consider here  $M=10$ and $q=1$.}
\label{LFxF_Culetu}
\end{figure}

From Eq.~\eqref{F}, we obain $r(F)$, which allows us to write ${\cal L}_F$ in terms of $F$, which is given by 
%
\begin{eqnarray}
&&    {\cal L}_F(F)=\exp\left\{-\frac{q^{2}}{8M^{2}}\left[e^{2}\text{Ei}\left(\frac{\sqrt[4]{F}q^{3/2}}{2^{3/4}M}-2\right)-5\text{Ei}\left(\frac{\sqrt[4]{F}q^{3/2}}{2^{3/4}M}\right)\right]
+\frac{\sqrt[4]{2F}q^{3/2}}{2M}
+\frac{\sqrt[4]{2F}q^{3/2}+M}{\sqrt{2F}Mq}\;
e^{\frac{\sqrt[4]{F}q^{3/2}}{2^{3/4}M}}\right\}\times
	\nonumber\\
&& \hspace{3cm}\times    \frac{M}{2F\kappa^{2}}
\left(\sqrt[4]{2F}q^{3/2}-8M\right)\left(\sqrt[4]{2F}q^{5/2}-4Mq\right)^{2}.\label{LF3_Culetu}
\end{eqnarray}
We present the behavior of Eq. \eqref{LF3_Culetu} in Fig. \ref{LFxF_Culetu}, which depicts ${\cal L}_F(F)$ as a non-linear bijective function with respect to $F$. 

\begin{figure}[h]
\centering
\includegraphics[scale=0.4]{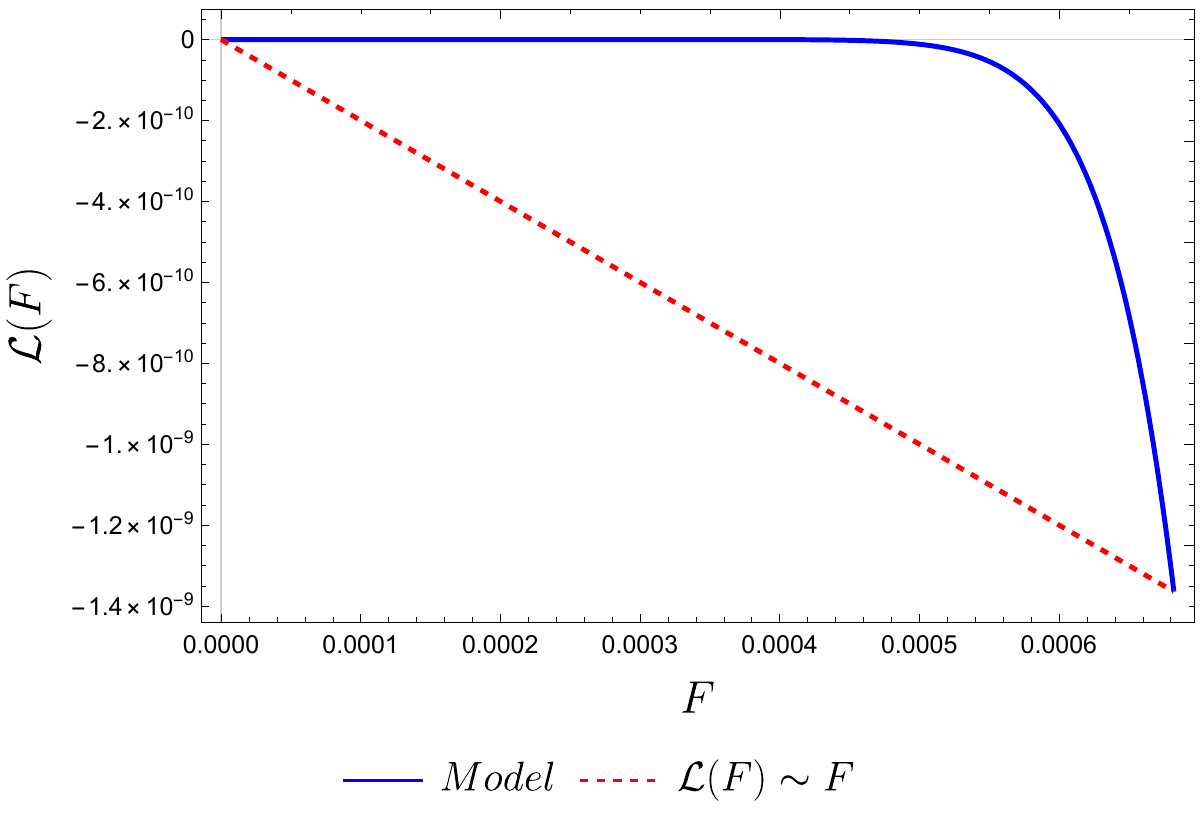}
\caption{Graphic representation of ${\cal L}(F)$, described by expression~\eqref{L_Culetu}. We consider here  $M=10$ and $q=1$. The blue curve represents the behavior of our model for $F\ll1$, while the dashed red line represents the linear case.} 
\label{LxF2_Culetu}
\end{figure}

For a more comprehensive understanding, we depict the asymptotic limits of the  Lagrangian \eqref{L_Culetu} in two distinct graphs. In Figure \ref{LxF2_Culetu}, we illustrate the behavior of ${\cal L} (F)$ for $F\ll1$. In this context, we compare our Culetu model with the linear case, depicted by the blue curve and the dashed red line, respectively. Meanwhile, we explore the behavior of ${\cal L} (F)$ for $F\gg1$, once again comparing it with the linear case.

\section{Summary and conclusion}\label{sec:conclusion}

In this work we investigated solutions for black holes and regular black holes in the newly proposed theory of $f(\mathbb{Q},\mathbb{B}_Q)$ gravity. This theory is a generalization of the $f(\mathbb{Q})$ theory, which describes gravitational effects through the non-metricity tensor $Q_{\beta\mu\nu}$. More specifically, we couple $f(\mathbb{Q},\mathbb{B}_Q)$ gravity with NLED and find generalizations of solutions for black holes and regular black holes. In particular, we thus further extend the class of regular solutions for models of the Bardeen and Culetu type black holes found in the literature. It is interesting to note that in the $f(\mathbb{Q})$ theory, in a coincident gauge, a constraint was found that renders the theory linear~\cite{Junior:2023qaq}, while in $f(\mathbb{Q},\mathbb{B}_Q)$ gravity, we gain a certain freedom which is not necessarily linear.

In the first model, the solution~\eqref{A1_BN} can be asymptotically Minkowskian if $b_0=-2$. However, it is also possible to choose values for $b_0$ that are smaller than $-2$, but in this case, the metric function~\eqref{A1_BN} remains asymptotically flat and the metric signature becomes ($-,+,-,-$).
This solution also generalizes GR with a term proportional to $1/r^4$. When $b_0=-2$ and $b_1=0$, we arrive at the Schwarzschild solution. However, if $b_1\neq0$, we get a different geometry that is still asymptotically flat. Therefore, it is interesting to study the physical properties of this solution, such as shadows and gravitational lensing, for example. Although the non-metricity scalar in Eq.~\eqref{Q_BN1} is regular, the combination $\mathbb{Q}-\mathbb{B}_Q$ is singular, rendering a singular spacetime geometry.

In the second model, described by the metric function~\eqref{A2_BN}, the latter diverges in the limit $r\to0$, while, as $r\to\infty$, $A(r)$ becomes asymptotically flat. If we choose $q_0=-2$, the metric function $A(r)$ becomes asymptotically Minkowskian. Although the non-metricity scalar is regular Eq.~\eqref{Q2_BN}, this does not guarantee that the spacetime solution is regular, as the combination of $\mathbb{Q}$ and $\mathbb{B}_Q$ for this model ensures that our solution is singular.

The equations of motion of the $f(\mathbb{Q},\mathbb{B}_Q )$ theory are quite complicated, coupled non-linear differential equations. Taking the vacuum as a possibility, there are two ways to proceed, one is to consider a functional form for the function $f(\mathbb{Q},\mathbb{B}_Q )$, and then integrate the equations to obtain the metric functions, first fixing a gauge. Another is to leave $f(\mathbb{Q},\mathbb{B}_Q )$ generic, fix a gauge, and integrate the equations to determine the metric functions and the functional form of $f(\mathbb{Q},\mathbb{B}_Q )$ together. These two paths seem impractical due to the complexity of the equations. Our strategy was to couple an NLED Lagrangian, leave the functions $\mathcal{L}$ and $\mathcal{L}_F$ free, solve the equations to determine these functions with respect to $f(\mathbb{Q},\mathbb{B}_Q )$ and the metric functions, and then set up the consistency equation such that $\mathcal{L}_F$ is the derivative of $\mathcal{L}$ with respect to $F$.  Then we define a model for the metric functions, integrate the consistency equation to determine $f(\mathbb{Q},\mathbb{B}_Q )$, and finally the functional form of $\mathcal{L}(F)$, if possible. This procedure has already been successfully applied to several other modified gravitational theories. \cite{Junior:2015dga,Junior:2015fya,Hollenstein:2008hp,Rodrigues:2015ayd,deSousaSilva:2018kkt,Capozziello:2019uvk}. The solutions obtained are generally of an NLED that does not become linear for the weak field limit, i.e. $F\ll1$, but this also happens for some solutions in general relativity, such as the first one suggested by Bardeen \cite{Ayon-Beato:2000mjt}.

In the third black hole model, we obtain the metric function described by Eq.~\eqref{A3_BN}. In the limit of $r\rightarrow0$, this metric function diverges; however, if in the limit of $r\rightarrow\infty$, we find that the metric function is asymptotically flat. However, if we choose $q_0=-2$, this metric function becomes asymptotically Minkowskian. Again, we find that the combination of the quantities $\mathbb{Q}$ and $\mathbb{B}_Q$  is singular, rendering a singular geometry. Analogously to regular black hole solutions that exhibit a regular spacetime, the derivative of the Lagrangian, ${\cal L}_F$, for both models of regular black holes is also distinct from its counterparts in GR.

Considering that the theory addressed in this work is of recent development, we see in this proposal a significant opportunity to explore new approaches. 
Our intention is to continue the study of $f(\mathbb{Q},\mathbb{B}_Q)$, exploring alternatives similar to those analysed in other theories, such as studies of black hole thermodynamics in $f(R)$ gravity~\cite{Akbar:2006mq,Bamba:2010kf,Zheng:2018fyn}, $f(T)$ gravity~\cite{Miao:2011ki,Bamba:2011pz,Karami:2012fu,Bamba:2012vg,Bamba:2012rv} and $f(T,\tilde{B})$ gravitational theories~\cite{Bahamonde:2016cul}, perturbation theory in $f(R)$~\cite{Carloni:2007yv, delaCruz-Dombriz:2008ium,Hojjati:2012rf}, $f(R,T)$~\cite{Alvarenga:2013syu,Sharif:2014ioa}, $f(T)$~\cite{Chen:2010va,Izumi:2012qj} and $f(\mathbb{Q})$ gravities~\cite{BeltranJimenez:2019tme,Khyllep:2022spx}, as well as the study of black hole shadows and the analysis of gravitational lensing in $f(R)$~\cite{Ruggiero:2007jr,Yang:2008wu,Nzioki:2010nj,Lubini:2011pc}, $f(R,T)$~\cite{Alhamzawi:2015eiv} and $f(T)$ gravity~\cite{Wu:2014wra}. Furthermore, we intend to explore new solutions of black holes and regular black holes in the context of nonlinear electrodynamics and with scalar field, through other classes of connections, i.e., beyond the coincident gauge, as presented and applied in~\cite{De:2023xua,Paliathanasis:2023kqs,Bahamonde:2022zgj}
These are some of the areas we intend to address in the future to deepen our understanding and contribute to the development of this new modified theory of gravity.


\begin{acknowledgments}
M. E. R. thanks CNPq for partial financial support.  This study was supported in part by the Coordenção de Aperfeioamento de Pessoal de Nível Superior - Brazil (CAPES) - Financial Code 001.  
FSNL acknowledges support from the Funda\c{c}\~{a}o para a Ci\^{e}ncia e a Tecnologia (FCT) Scientific Employment Stimulus contract with reference CEECINST/00032/2018, and funding through the research grants UIDB/04434/2020,  UIDP/04434/2020 and PTDC/FIS-AST/0054/2021.
\end{acknowledgments}

\end{document}